\documentclass[%
 reprint,
 amsmath,amssymb,
prx,
]{revtex4-2}

\usepackage{bm}
\usepackage{amssymb}
\usepackage{amstext}
\usepackage{amsmath}
\usepackage{mathrsfs}
\usepackage{graphicx}
\usepackage{color}
\usepackage{amsthm}
\usepackage{float}
\usepackage{array}

\usepackage{listings}
\usepackage{booktabs}
\usepackage{algorithm2e}

\begin{document}
\preprint{APS/123-QED}
\title{Bridging the information and dynamics attributes of neural activities}
\thanks{Correspondence should be addressed to Y.T and P.S.}%
\author{Yang Tian}
\email{tiany20@mails.tsinghua.edu.cn}
 \altaffiliation[]{Department of Psychology \& Tsinghua Brain and Intelligence Lab, Tsinghua University, Beijing, 100084, China.}
  \author{Guoqi Li}%
 \email{liguoqi@mail.tsinghua.edu.cn}
 \altaffiliation[]{Department of Precision Instrumentation \& Center
for Brain Inspired Computing Research, Tsinghua University, Beijing 100084, China.}
\author{Pei Sun}%
 \email{peisun@tsinghua.edu.cn}
  \altaffiliation[]{Department of Psychology \& Tsinghua Brain and Intelligence Lab, Tsinghua University, Beijing, 100084, China.}




\begin{abstract}
The brain works as a dynamic system to process information. Various challenges remain in understanding the connection between information and dynamics attributes in the brain. The present research pursues exploring how the characteristics of neural information functions are linked to neural dynamics. We attempt to bridge dynamics (e.g., Kolmogorov-Sinai entropy) and information (e.g., mutual information and Fisher information) metrics on the stimulus-triggered stochastic dynamics in neural populations. On the one hand, our unified analysis identifies various essential features of the information-processing-related neural dynamics. We discover spatiotemporal differences in the dynamic randomness and chaotic degrees of neural dynamics during neural information processing. On the other hand, our framework reveals the fundamental role of neural dynamics in shaping neural information processing. The neural dynamics creates an oppositely directed variation of encoding and decoding properties under specific conditions, and it determines the neural representation of stimulus distribution. Overall, our findings demonstrate a potential direction to explain the emergence of neural information processing from neural dynamics and help understand the intrinsic connections between the informational and the physical brain.
\end{abstract}

\maketitle
\section{Introduction}
Understanding how the brain works is one of the most frontier directions at the intersection of biology and physics \cite{collell2015brain}. Till now, substantial progress has demonstrated that the brain can be treated as a dynamic system that processes information \cite{dayan2001theoretical}. The brain is frequently driven out of equilibrium by external stimuli \cite{shah2004neural,vogels2005neural,deco2012neural} or internal events \cite{foster2016spontaneous,galan2008network} and creates multifarious dynamics \cite{coombes2010large,chialvo2010emergent,lumer1997neural1,lumer1997neural2,martinello2017neutral}. In the meantime, when the dynamics is stimulus-triggered, the stimulus information is coded \cite{borst1999information} and memorized \cite{amit1987information} by the brain to support cognitive functions \cite{purves2008cognitive,butts2007temporal,laurent1994encoding}, making the brain an information system as well \cite{averbeck2006neural}. Research into such dual attributes of the brain features a long history. Extensive connections between the dynamics and information attributes have been discovered in the brain \cite{harth1970brain,ermentrout2007relating,murray2017stable,vogels2005neural}, indicating that the cognitive functions that process external information (or referred to as information functions) are essentially rooted in neural dynamics \cite{cessac2010overview,brenner2000synergy,schroeder2010dynamics}.

Although the analyses on neural dynamics \cite{coombes2010large,chialvo2010emergent,lumer1997neural1,lumer1997neural2,martinello2017neutral} and neural information attributes \cite{borst1999information,amit1987information,skaggs1993information,linsker1990perceptual,purves2008cognitive,butts2007temporal,laurent1994encoding,averbeck2006neural} have seen substantial progress respectively, it remains an open question on how dynamics and information feature such fundamental connections in the brain. The pursuit of an appropriate answer to this question faces various challenges, among which, a critical one is the shortage of a practical analysis framework that unifies the dynamics \cite{katok1997introduction} and information \cite{khinchin2013mathematical} quantities in neural activities. Regarding neural activity characterization, a dilemma exists that the stochastic models \cite{paninski2004maximum,heeger2000poisson,barbieri2001construction,burkitt2006review} surpass deterministic models \cite{hodgkin1952quantitative,stein1965theoretical,izhikevich2003simple,gerstner2002spiking,tsodyks1998neural} in supporting information-theoretical metrics, but these stochastic approaches are weak in defining the dynamics related to inter-neuron interactions and neural tuning properties (i.e., the response selectivity to stimuli). Meanwhile, another challenge arises from the lack of an applicable metric of the neural dynamics involved in information processing. In comparison with the information-theoretical metrics rooted in probabilistic frameworks (e.g., mutual information \cite{borst1999information}), mainstream dynamics-theoretical metrics in experimental \cite{grossberg1987neural,grossberg2011neural,seelig2015neural} and theoretical \cite{buesing2011neural,herz2006modeling,dror2000chaos,amari1977dynamics,kishimoto1979existence,rougier2006dynamic,hutt2003pattern,martinello2017neutral,montbrio2015macroscopic,engelken2020lyapunov,sompolinsky1988chaos} studies are mainly built on non-probabilistic dynamics theories (e.g., Lyapunov spectra \cite{engelken2020lyapunov,sompolinsky1988chaos}), impeding an analytical unification with information quantities. 

\begin{figure*}[!t]
\includegraphics[width=2\columnwidth]{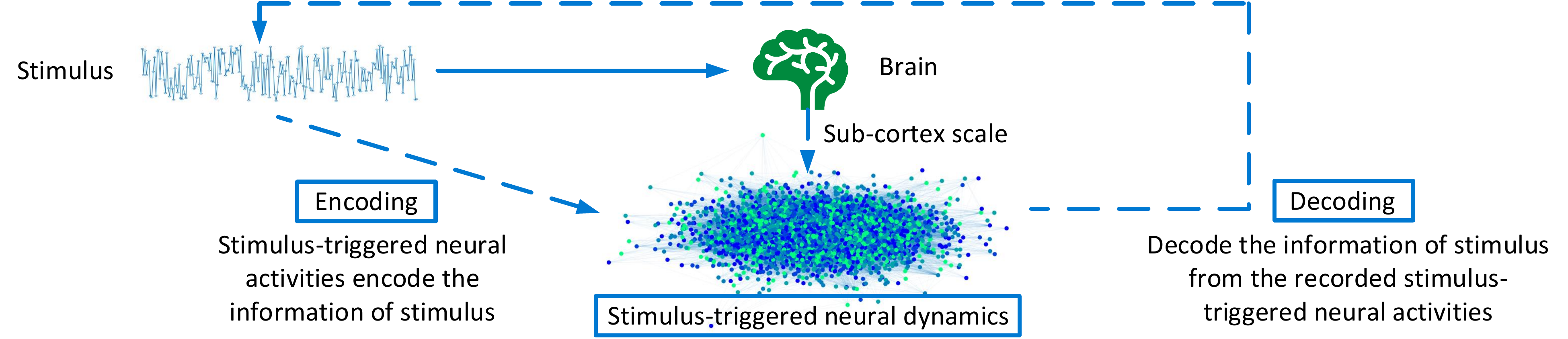}
\caption{The connections between neural information functions and neural dynamics.} 
\end{figure*}

The present research pursues to glance at the intrinsic relations between information and dynamics in the brain. To build an operable framework, we concentrate on the elementary neural information functions from encoding and decoding perspectives and the stimulus-triggered neural dynamics in neural populations (see Fig. 1). We attempt to implement a unified analysis of these elements and explore the emergence of the characteristics of neural information functions from the dynamics of neural ensembles. Technically, our research may contribute to developing a possible description of stimulus-triggered neural activities in neural populations, supporting the analytic measurement of dynamics and information quantities during neural information processing. Theoretically, the significance of our pursuit lies in the possibility for the analysis to explore the fundamental connections between the physical (dynamics aspect) and informational (information aspect) brain \cite{collell2015brain}. These connections may reveal a potential direction to study why the complex and remarkable characteristics of neural information processing and cognition can naturally emerge in the brain, a system of the neurons that only have elementary functions.

The paper is organized as follows. In Sec. \ref{Sec1}, we introduce a mathematical characterization of the stimulus-triggered neural activities during information processing, whose foundation has been established in our previous research \cite{tian2021characteristics}. Based on this characterization, the dynamic randomness and chaotic degree of stimulus-triggered neural activities are quantified in terms of Kolmogorov-Sinai entropy (Sec. \ref{Sec2}). After reviewing the quantification of neural information function attributes (e.g., encoding and decoding efficiency), we explore the substantive characteristics of information-processing-related neural dynamics and analyze the emergence of neural information function attributes from neural dynamics. Various potential connections between the information and dynamics attributes of neural activities are observed (Sec. \ref{Sec3}). In Sec. \ref{Sec4}, we provide an integrated and multi-scale perspective for our theoretical framework and computational findings. We attempt to verify our discoveries' validity and generalization ability by relating them with existing experiment-validated studies or proposing mechanistic insights into why they arise. Finally, we discuss several potential directions and remaining challenges for future explorations. While we concentrate on physical pictures and neuroscience backgrounds throughout the paper, one can find the systematic description of all mathematical implementations in appendixes. 

\section{Stimulus-triggered neural activities}\label{Sec1}
\subsection{Neural population description}

We begin with a neural population $\mathcal{N}\left(\mathcal{V},\mathcal{E}\right)$, where $\mathcal{V}$ is the neuron set and $\mathcal{E}$ is the synapse set. The synaptic connection strength is defined by an adjacent matrix $\mathcal{C}$ ($\mathcal{C}_{ij}\in\left[-1,1\right]$). We randomize $\mathcal{V}$, $\mathcal{E}$, and $\mathcal{C}$ for generality (see appendix \ref{1.1M}). In real neural populations, the stimuli will not be simultaneously received by all neurons. Stimulus information experiences a complex diffusion process among neurons, creating time differences for neurons to receive stimuli. Therefore, it is biologically reasonable to classify these neurons into input neurons (receive inputs directly) and intermediary neurons (triggered by their pre-synaptic neurons and process inputs indirectly) (see Fig. 2(a) for an example). 

\begin{figure*}[!t]
\includegraphics[width=2\columnwidth]{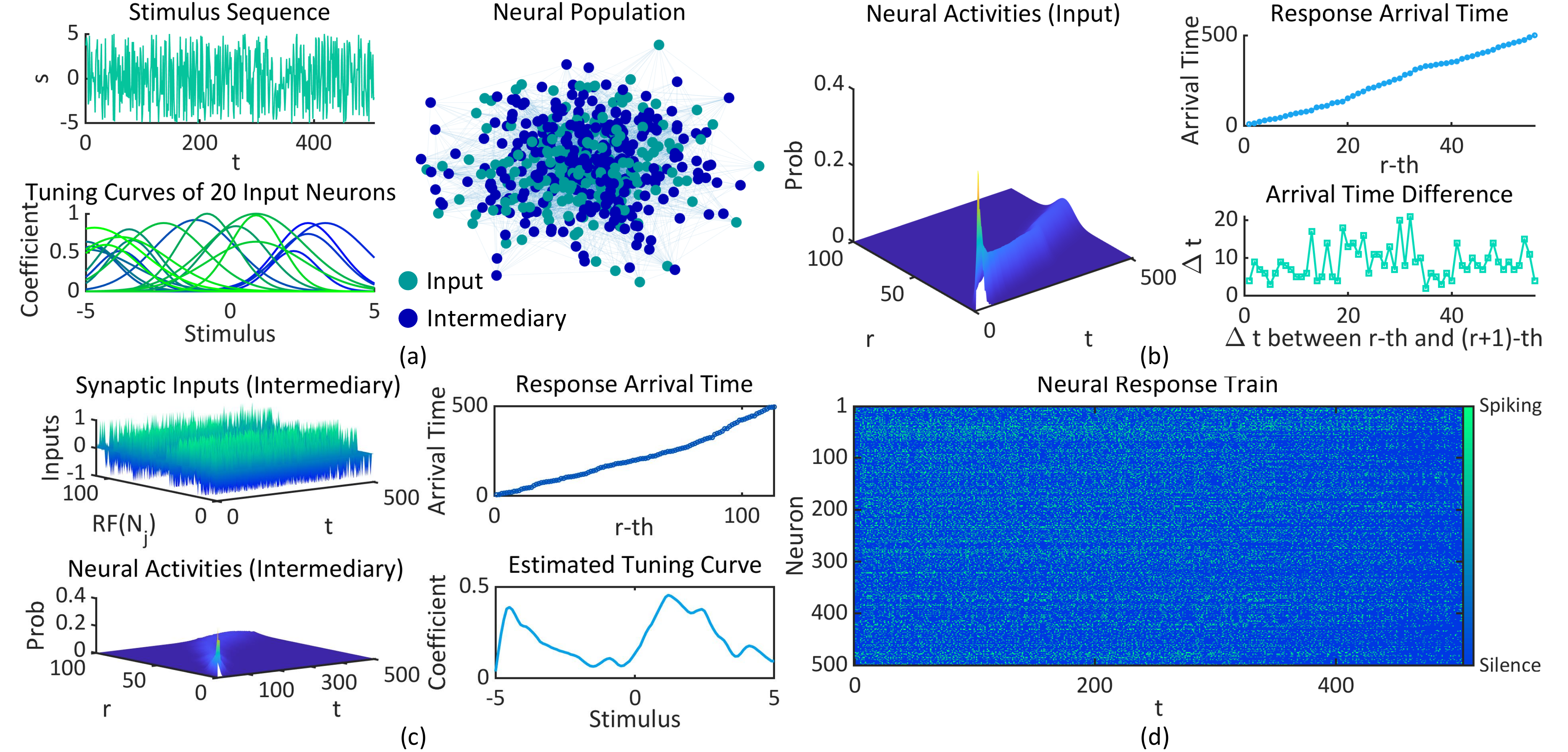}
    \caption{Stimulus-triggered neural activity characterization. \textbf{(a)} A population with 500 neurons (the ratio of input neurons to intermediary neurons is $3:2$), a random stimulus sequence $\mathcal{S}^{\prime}=\lbrace s_{k}\mid s_{k}\in\left[-5,5\right]\rbrace$ in time interval $\left[0,500\right]$, and 20 examples of input neuron tuning curves with $\widehat{r}\in\left[0.5,1\right]$, $\widehat{s}\in\left[-5,5\right]$, and $\sigma\in\left[\frac{5}{6},\frac{5}{3}\right]$. \textbf{(b)} The stochastic process of the neural activities of a randomly picked input neuron, the predicted neural response arrival sequence, and the time difference between each two predicted responses. \textbf{(c)} The pre-synaptic inputs, the estimated stochastic process of neural activities ($h=100$), the observed neural response sequence, and the estimated tuning curve (for visualization, it is smoothed from the raw data utilizing the Savitzky-Golay filter \cite{press1990savitzky}) of the intermediary neuron. \textbf{(d)} The estimated neural response train.} 
\end{figure*}

\subsection{Neural activities of input neurons}
Input neurons process stimuli directly, their activity profiles are mainly determined by their neural tuning properties. A standard characterization for the tuning property is the tuning curve \cite{dayan2001theoretical,butts2006tuning}, where the stimulus at the peak evokes the highest response rate \cite{butts2006tuning}. In our research, each input neuron $N_{i}$ has a bell-shaped tuning curve $\mathcal{G}_{i}\left(s\right)$ with a maximum response coefficient $\widehat{R}_{i}$, a preferred stimulus $\widehat{s}_{i}$, and a curve width $\sigma_{i}$ (see (\ref{TuningCurve}) in appendix \ref{1.2M} and Fig. 2(a)). Assuming a randomized stimulus sequence $\mathcal{S}^{\prime}$ occurs in a time interval $\left[0,t^{\prime\prime}\right]$, we implement the neural activity characterization as the probability $\mathcal{P}_{i}\left(r\mid \mathcal{S}^{\prime},0,t\right)$ for the cumulative neural response count of input neuron $N_{i}$ to reach a specific quantity $r$ at any moment $t$ in $\left[0,t^{\prime\prime}\right]$. This probability can be efficiently approximated by the Poisson process \cite{dayan2001theoretical}. Specifically, we use the tuning curve $\mathcal{G}_{i}\left(s\right)$ as the intensity function of Poisson process to describe the response selectivity of $N_{i}$ on $\mathcal{S}^{\prime}$. Such definition makes the Poisson process non-homogeneous, realizing a time-varying neural activity intensity controlled by neural tuning properties 
\begin{align}
\mathcal{P}_{i}\left(r\mid \mathcal{S}^{\prime},0,t\right)&=\dfrac{\left[\Lambda_{i}\left(0,t\right)\right]^{r}}{r!}\exp\left(-\Lambda_{i}\left(0,t\right)\right),\label{EQ1}\\
\Lambda_{i}\left(0,t\right)&=\int_{0}^{t}\mathcal{G}_{i}\left(\mathcal{S}^{\prime}\left(m\right)\right)dm \label{EQ2}
\end{align}
(please see Fig. 2(b) and (\ref{MasterEq}-\ref{TDF}) in appendix \ref{1.2M}). 

Given the probability distribution of the Poisson process, we can generate possible neural activities by predicting the arrival time of each neural response of $N_{i}$. We implement the prediction with the maximum probability method (MP). This method estimates the arrival time of $r$-th neural response as the moment $\widehat{t}_{r}$ when the response frequency is most likely to reach $r$ 
\begin{align}
\widehat{t}_{r}=\operatorname*{argmax}_{t}\left(\mathcal{P}_{i}\left(r\mid \mathcal{S}^{\prime},0,t\right)\right)\label{EQ3}
\end{align}
(see (\ref{MPM}) in appendix \ref{1.2M}). Fig. 2(b) illustrates an example of the predicted arrival time sequence and the time difference between each two neural responses. With the predicted sequence $\mathcal{T}_{i}=\lbrace\widehat{t}_{r}\rbrace_{r\in\mathbb{N}^{+}}$, we can mark every neural response of $N_{i}$ in the time line by $\mathcal{R}_{i}\left(t\right)$, where $\mathcal{R}_{i}\left(t\right)=1$ stands for response and $\mathcal{R}_{i}\left(t\right)=0$ stands for no response (see (\ref{MarkerofNR}) in appendix \ref{1.2M}).

\subsection{Neural activities of intermediary neurons}
Intermediary neurons are driven by their pre-synaptic neurons rather than direct stimulus inputs. It is unreasonable to limit their activity profiles by pre-setting their tuning curves. Obviously, their activities are significantly affected by the network dynamics, leading to obstacles for a priori stochastic characterization.

We develop a two-step statistical approach to overcome these obstacles. For an intermediary neuron $N_{j}$ that features a receptive field $\mathcal{RF}\left(N_{j}\right)$ (the set of pre-synaptic neurons), the first step is to predict the neural activity arrival sequence of each neuron $N_{k}$ in $\mathcal{RF}\left(N_{j}\right)$ and define the synaptic inputs given from $N_{k}$ to $N_{j}$ as $\mathcal{R}_{k}\left(t\right)\mathcal{C}_{kj}$ (see Fig. 2(c)). By summing these inputs over all neurons in $\mathcal{RF}\left(N_{j}\right)$, we can obtain the total synaptic input $\Psi_{j}$ of neuron $N_{j}$. Then we characterize the neural response of $N_{j}$ following
\begin{align}
\widehat{\mathcal{R}}_{j}\left(t\right)=\upsilon\left[\int_{0}^{t}\Big(\Psi_{j}\left(x\right)+\Omega_{j}\left(x\right)\Big)dx-\Upsilon_{j}\left(t\right)\right], \label{EQ4}
\end{align}
where $\upsilon\left(\cdot\right)$ denotes the unit step function. This integrate-and-fire response mechanism has a dissipation term $\Omega$ (e.g., the leaky term in the leaky integrate-and-fire neuron \cite{burkitt2006review}) and a neural response threshold $\Upsilon$ (e.g., spiking threshold \cite{burkitt2006review}), implementing that $N_{j}$ emits a response only if $\text{cumulative synaptic inputs}-\text{dissipation}>\text{threshold}$ (see (\ref{PostNeuralResponse}-\ref{Threshold}) in appendix \ref{M1.3}). The concrete examples of this mechanism can be seen in existing deterministic models \cite{hodgkin1952quantitative,stein1965theoretical,izhikevich2003simple,gerstner2002spiking,tsodyks1998neural}. The second step is to treat the generated neural activities of $N_{j}$ as observed samples and repeat the generation with $\mathcal{S}^{\prime}$ for $h$ times. This repeated sampling supports a maximum likelihood estimation for the intensity function of the neural activities of $N_{j}$, constructing an observed distribution of the Poisson process
\begin{align}
\widehat{\mathcal{P}}_{j}\left(r\mid \mathcal{S}^{\prime},0,t\right)&=\dfrac{\left[\widehat{\Lambda}_{j}\left(0,t\right)\right]^{r}}{r!}\exp\left(-\widehat{\Lambda}_{j}\left(0,t\right)\right),\label{EQ5}\\
\widehat{\Lambda}_{j}\left(0,t\right)&=\int_{0}^{t}\frac{1}{h}\sum_{a\in \mathbb{Z}\cap\left[1,h\right]}\widehat{\mathcal{R}}_{j,a}\left(m\right)dm. \label{EQ6}
\end{align}
(\ref{EQ5}) can be further applied to estimate the neural response arrival sequence and the neural tuning curve of $N_{j}$ (see Fig. 2(c) and (\ref{InterveingIntensityF}-\ref{ObTC}) in appendix \ref{M1.3}). Such a two-step approach takes the advantages of deterministic models in describing inter-neuron interactions (see (\ref{EQ4})). By estimating Poisson processes based on the generated activities in (\ref{EQ4}), the difficulties underlying a direct probabilistic description of network dynamics are avoided.

\begin{figure*}[!t]
\includegraphics[width=2\columnwidth]{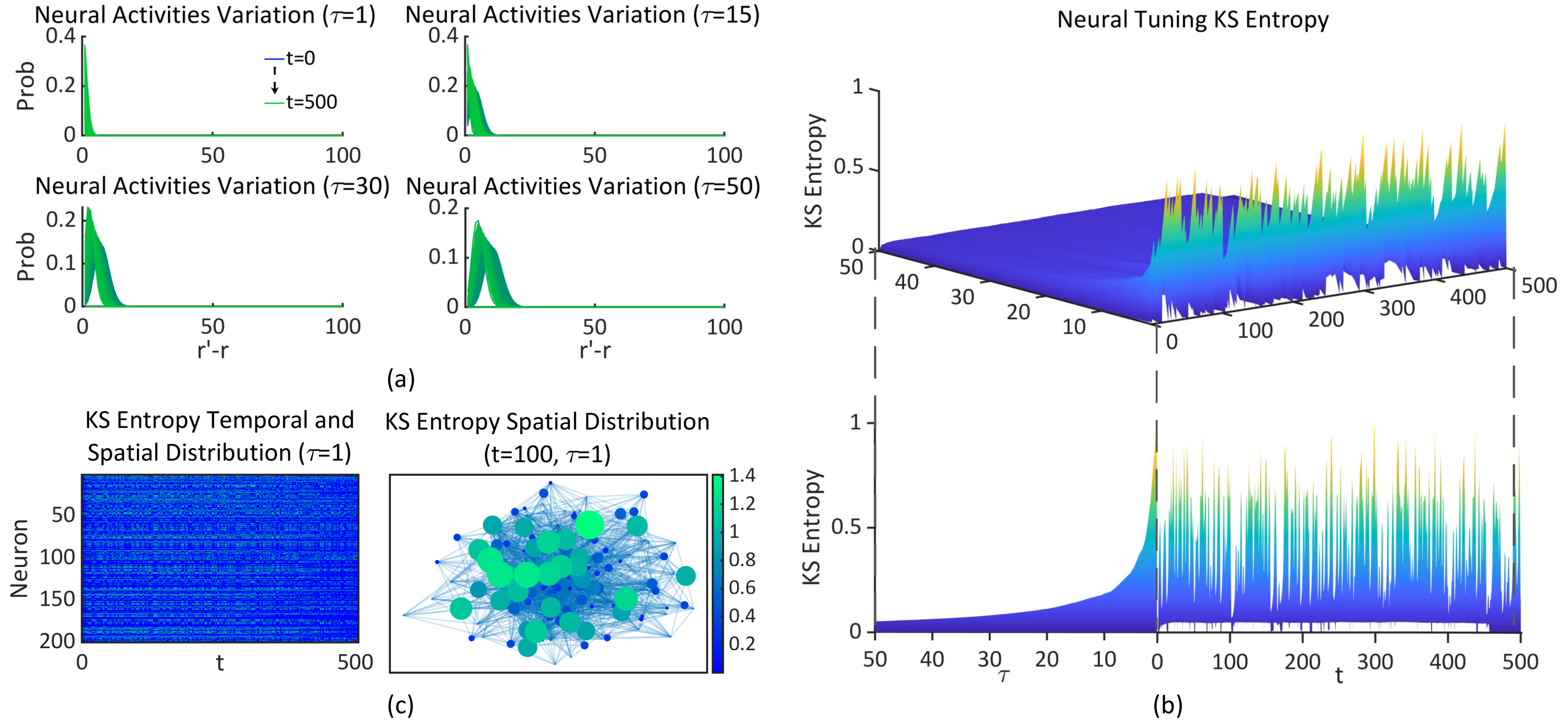}
\caption{Neural dynamics measurement. We randomize a stimulus sequence $\mathcal{S}^{\prime}=\lbrace s_{k}\mid s_{k}\in\left[-5,5\right]\rbrace$ in $\left[0,500\right]$ and a neural population with 200 neurons (the ratio of input neurons to intermediary neurons is $3:2$). \textbf{(a)} The probability distribution $\mathcal{P}_{n}^{\heartsuit}\left(r^{\prime}-r\mid\mathcal{S}^{\prime},t,t+\tau\right)$ of an arbitrary neuron $N_{n}$ with different $\tau$. \textbf{(b)} The entropy $\mathcal{H}_{KS}$ of neuron $N_{n}$. \textbf{(c)} The temporal and spatial distribution of $\mathcal{H}_{KS}$ ($\tau=1$), and the spatial distribution of $\mathcal{H}_{KS}$ ($t=100$, $\tau=1$) in the neural population.} 
\end{figure*}

For a summary, we define $\mathcal{P}_{n}^{\heartsuit}\left(r\mid \mathcal{S}^{\prime},0,t\right)$ to describe the neural activities of neuron $N_{n}$, where $\heartsuit$ denotes the neuron type ($\heartsuit=1$ for the input neuron and $\heartsuit=0$ for the intermediary neuron). The algorithm in appendix \ref{1.4M} depicts our framework. Fig. 2(d) illustrates the estimated neural response train by this framework. 

\section{Dynamics in stimulus-triggered neural activities}\label{Sec2}
\subsection{Neural tuning Kolmogorov-Sinai entropy}
 A remaining challenge is to develop a dynamics-theoretical metric for stimulus-triggered neural activities. Although a natural choice is Kolmogorov-Sinai entropy, which can characterize the randomness of a dynamic system forward in time \cite{lecomte2007thermodynamic}, the implementation of this entropy in neural dynamics remains non-trivial. To simplify the calculation of the Kolmogorov-Sinai entropy, we reformulate the proposed Poisson processes as continuous Markov chains 
 \begin{align}
&\frac{\partial}{\partial t}\mathcal{P}_{n}^{\heartsuit}\left(r\mid \mathcal{S}^{\prime},t,t+\tau\right)=\notag\\&\sum_{r^{\prime}\leq r}\mathbb{W}_{r^{\prime}r}\left(\mathcal{S}^{\prime},t,t+\tau\right)-\sum_{r^{\prime}> r}\mathbb{W}_{rr^{\prime}}\left(\mathcal{S}^{\prime},t,t+\tau\right), \label{EQ7}
\end{align}
where parameter $\tau\geq0$ measures the interval length of variation. We define that $\mathbb{W}_{ij}\left(\mathcal{S}^{\prime},t,t+\tau\right)=\mathcal{W}_{ij}\left(\mathcal{S}^{\prime},t,t+\tau\right)\mathcal{P}_{n}^{\heartsuit}\left(i\mid \mathcal{S}^{\prime},0,t\right)$, in which $\mathbb{W}$ denotes the transition probability matrix (one can see (\ref{TransM}-\ref{MasterEqofMarkov2}) in appendix \ref{2.1M} for details). Then, a new Kolmogorov-Sinai entropy $\mathcal{H}_{KS}\left(N_{n},\mathcal{S}^{\prime},t,t+\tau\right)$ (referred to as neural tuning Kolmogorov-Sinai entropy)
\begin{align}
&\mathcal{H}_{KS}\left(N_{n},\mathcal{S}^{\prime},t,t+\tau\right)=-\frac{1}{\tau}\sum_{r}\sum_{r^{\prime}> r}\mathcal{P}_{n}^{\heartsuit}\left(r\mid\mathcal{S}^{\prime},0,t\right)\notag\\
&\mathcal{W}_{rr^{\prime}}\left(\mathcal{S}^{\prime},t,t+\tau\right)\ln\mathcal{W}_{rr^{\prime}}\left(\mathcal{S}^{\prime},t,t+\tau\right)\label{EQ8}
\end{align}
is proposed as a metric of the dynamic randomness of the activities of neuron $N_{n}$ in a time interval $\left[t,t+\tau\right]$ (see (\ref{ClassicalDefinitionKS}-\ref{NKSE}) in appendix \ref{2.2M}). Note that the interval length $\tau$ should not be too small since any biological variation takes time in the neural system. Our research considers the cases where $\tau\geq 1$. 

\begin{figure*}[!t]
\includegraphics[width=2\columnwidth]{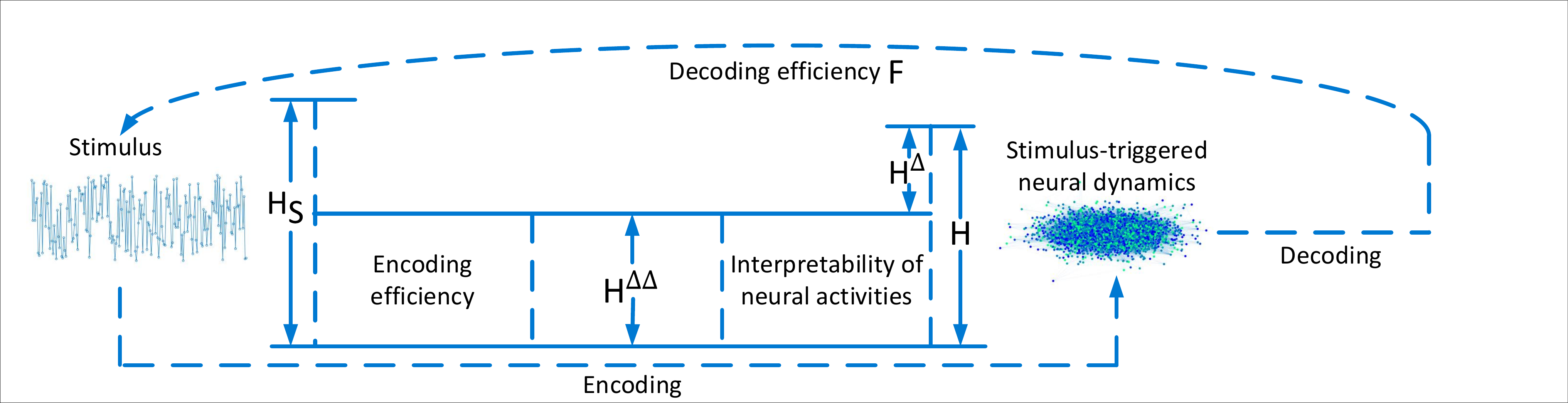}
\caption{The relations between different neural information function properties.}
\end{figure*}  

Fig. 3(a) shows that the maximum possible variation amplitude of neural activities (the maximum change $r^{\prime}-r$ with non-zero $\mathcal{P}_{n}^{\heartsuit}\left(r^{\prime}-r\mid\mathcal{S}^{\prime},t,t+\tau\right)$) is positively correlated with $\tau$. In Fig. 3(b), we find that $\mathcal{H}_{KS}\left(N_{n},\mathcal{S}^{\prime},t,t+\tau\right)$ is larger when $\tau$ is relatively small. Together, we can know that while short-term neural activities have relatively small amplitudes of variations, the dynamic randomness of those variations are more complex. In comparison, long-term variations feature larger amplitudes, but the variation tendency is relatively stable. Moreover, Fig. 3(c) shows the temporal and spatial distribution of $\mathcal{H}_{KS}$ with $\tau=1$ (measures the short-term dynamic randomness) on the population scale, and a spatial distribution of $\mathcal{H}_{KS}$ with $\tau=1$ and $t=100$.
\subsection{Chaos in neural activities}
An important property of the Kolmogorov-Sinai entropy is that it is bound by the summation of all the positive Lyapunov exponents of the dynamic system (please see Pesin identity \cite{eckmann1985ergodic} and further see Ruelle inequality \cite{ruelle1978inequality} for a generalization). This mathematical relation bridges between our metric and the Lyapunov spectra analysis \cite{engelken2020lyapunov,sompolinsky1988chaos}. Each Lyapunov exponent characterizes the separation or convergence rate of different infinitesimally close trajectories in the phase space of a dynamic system, and a positive Lyapunov exponent reflects the existence of chaos. The existence of such a property suggests that the neural activities are chaotic when the corresponding $\mathcal{H}_{KS}$ is positive, and the chaos will be intensified when $\mathcal{H}_{KS}$ increases (see appendix \ref{2.3M}).

\section{Bridge the information and dynamics attributes}\label{Sec3}
\subsection{Information-theoretical metrics reformulation}
Our research concentrates on the relations between the stimulus-triggered neural dynamics and the elementary neural information functions (studied from the aspects of encoding and decoding). Our proposed neural activity characterization above supports the analytical calculation of encoding-efficiency-related and decoding-efficiency-related metrics. 

For neural encoding, we concentrate on the widely-used metrics such as total response entropy $\mathcal{H}$ (TE), noise entropy $\mathcal{H}^{\vartriangle}$ (NE) and mutual information $\mathcal{H}^{\vartriangle\vartriangle}$ (MI). They respectively measure the total variability of neural responses to stimuli, the inexplicable part of the variability by stimuli, and the explicable part of the variability by stimuli. Their classic definitions are summarized in \cite{dayan2001theoretical}. In the present research, they are reformulated to be time-dependent to fit dynamic situations (see (\ref{TRE}-\ref{MI}) in appendix \ref{3.1M}). At each moment $t$, we can directly measure $\mathcal{H}\left(N_{n},t\right)$, $\mathcal{H}^{\vartriangle}\left(N_{n},t\right)$, and $\mathcal{H}^{\vartriangle\vartriangle}\left(N_{n},t\right)$ for every neuron $N_{n}$. Meanwhile, we can measure the entropy of stimulus sequence (ES) as $\mathcal{H}_{\mathcal{S}}\left(\mathcal{S}^{\prime},t\right)$ (see (\ref{EofS}) in appendix \ref{3.1M}). Moreover, we can use $\frac{\mathcal{H}^{\vartriangle\vartriangle}}{\mathcal{H}}$ (MI/TE) to measure the interpretability of neural activities based on stimuli, and use $\frac{\mathcal{H}^{\vartriangle\vartriangle}}{\mathcal{H}_{\mathcal{S}}}$ (MI/ES) to measure the encoding efficiency for stimuli based on neural activities.

The neural decoding efficiency is measured by Fisher information (FI) \cite{dayan2001theoretical,yarrow2012fisher}. In the present research, the measurement of the time-dependent Fisher information $\mathcal{F}\left(N_{n},t\right)$ of each neuron $N_{n}$ is proposed by (\ref{FisherInformation}) in appendix \ref{3.2M}. Cram$\acute{\text{e}}$r-Rao bound suggests that Fisher information limits the accuracy with which any decoding technique can estimate about the target stimulus parameter based on neural activities \cite{dayan2001theoretical}. Thus, the time-dependent Fisher information $\mathcal{F}\left(N_{n},t\right)$ acts as the lower bound of the variance of any decoding technique applied on neuron $N_{n}$. Any decoding scheme that reaches this variance bound is optimal \cite{dayan2001theoretical}. 

Fig. 4 depicts the discussed relations between these information-theoretical metrics. Within such a frame, our unified analysis obtains four main findings.

\begin{figure*}[!t]
\includegraphics[width=2\columnwidth]{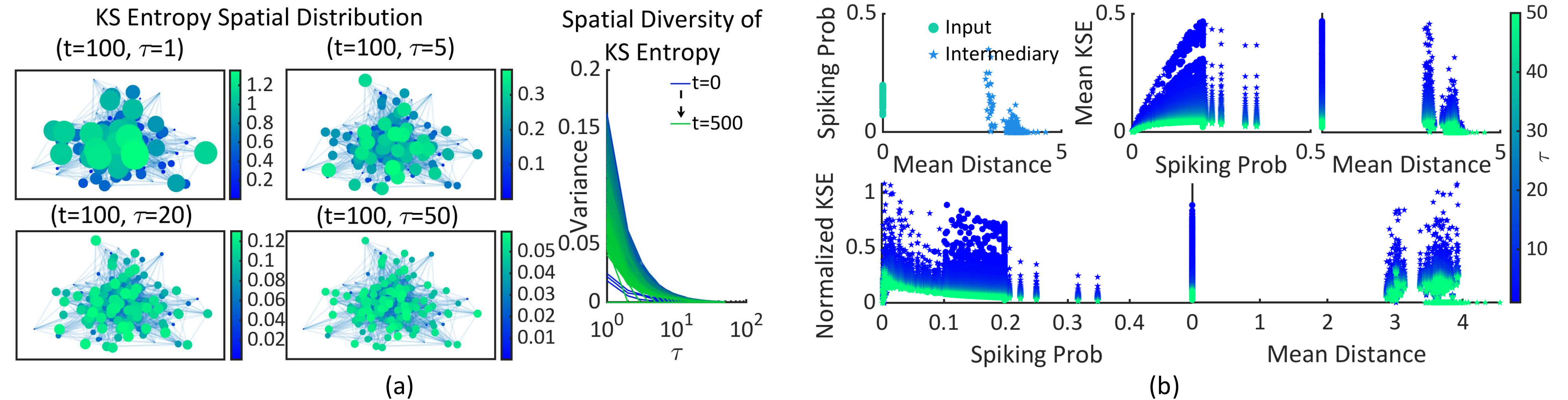}
\caption{Two findings of the stimulus-triggered neural dynamics. The experiment is based on a stimulus sequence $\mathcal{S}^{\prime}=\lbrace s_{k}\mid s_{k}\in\left[-5,5\right]\rbrace$ in $\left[0,500\right]$ and a neural population with 500 neurons (the ratio of input neurons to intermediary neurons is $1:1$). \textbf{(a)} We illustrate 4 instances of the spatial distribution of $\mathcal{H}_{KS}$ with $t=100$ and different $\tau$ (left and middle). Meanwhile, a quantification of the spatial diversity of $\mathcal{H}_{KS}$ is implemented utilizing the variance $\operatorname{Var}\left(\mathcal{H}_{KS}\right)$ among all neurons (right). \textbf{(b)} The mean distance between each neuron $N_{n}$ and all input neurons is calculated (this distance is set as 0 when $N_{n}$ is an input neuron). One can see that the spiking probability (spiking frequency) of each neuron reduces with the increase of distance (upper line, left). Meanwhile, the mean $\mathcal{H}_{KS}$ values (averaged through the time interval) of all neurons increase along with their spiking probability quantities (upper line, middle). Therefore, the mean $\mathcal{H}_{KS}$ values decrease as the distance increases (upper line, right). Conversely, one can see that the normalized $\mathcal{H}_{KS}$ values decline when the spiking probability increases (bottom line, left). Compared with the mean $\mathcal{H}_{KS}$, the normalized $\mathcal{H}_{KS}$ features an opposite trend, decreasing along with the mean distance (bottom line, right). Here the colors of all data points scale depending on the corresponding $\tau$.}
\end{figure*}

\subsection{Finding 1: The difference of dynamic randomness between the short-term and long-term variations in neural activities}

 As suggested above, compared with the long-term (large $\tau$) variations of neural activities, the short-term (small $\tau$) variations have small amplitudes but much more complex dynamic randomness (Fig. 3(a-b)). In Fig. 5(a), this finding is further verified under more general conditions. Based on the illustrated instances and statistical results, we find that the dynamic randomness measured by $\mathcal{H}_{KS}$ of each neuron, and the diversity of dynamic randomness between neurons, are negatively correlated with $\tau\in\left[1,50\right]$ (the diversity of dynamic randomness is quantified by the variance $\operatorname{Var}\left(\mathcal{H}_{KS}\right)$ among neurons). Such finding indicates that the short-term neural activities have smaller variation amplitudes but more complex dynamic randomness, implying larger inter-neuron diversity in dynamic randomness. Opposite characteristics are featured by long-term neural activities, namely, larger variation amplitudes but more stable trends, and the activities of all neurons tend to be more homogeneous.
 
\subsection{Finding 2: The uneven spatial distribution of chaos in neural populations}
As shown in Fig. 3(c), the spatial distribution of Kolmogorov-Sinai entropy $\mathcal{H}_{KS}$ in the neural population is uneven. The dynamic randomness varies between neurons (no matter at any specific moment or through the time interval). To offer a more solid verification, we set another random neural population with 500 neurons (the ratio of input neurons to intermediary neurons is $1:1$). We treat all input neurons as the input ports of this population and calculate the mean minimum distance (the shortest path length on the graph) between each neuron $N_{n}$ and these input ports (the distance is set as $0$ when $N_{n}$ happens to be an input neuron). An intermediary neuron with a larger distance is treated as in a deeper layer, defining a direction from shallow to deep layers. 

In the upper panel of Fig. 5(b), we analyze the dynamic randomness quantified by the mean $\mathcal{H}_{KS}$ of each neuron (averaged through the time interval $\left[0,500\right]$). We find that the neural spiking probability (approximated by the spiking frequency in the interval $\left[0,500\right]$) declines along with the mean distance to input neurons. Because of the positive correlation between the mean $\mathcal{H}_{KS}$ and the spiking probability, the mean $\mathcal{H}_{KS}$ decreases from shallow layers to deep layers. In the bottom panel of Fig. 5(b), we concentrate on the dynamic randomness of the neural responses to stimuli. For each neuron, we define the mean $\mathcal{H}_{KS}$ only using the raw data of $\mathcal{H}_{KS}$ while it spikes. Then the newly calculated mean $\mathcal{H}_{KS}$ of each neuron is normalized by dividing the corresponding spiking probability. This normalized $\mathcal{H}_{KS}$ reflects the dynamic randomness when a neuron responds to stimuli (i.e., generates spikes) . The normalization prevents this metric from increasing with the spiking probability sharply. Thus, we can see that the normalized $\mathcal{H}_{KS}$ relatively increases from shallow layers to deep layers.

Given the connection between $\mathcal{H}_{KS}$ and chaos, a positive $\mathcal{H}_{KS}$ suggests chaos in neural activities, and the chaos will be intensified with a larger $\mathcal{H}_{KS}$. Together, we conclude that the spatial distribution of chaos in neural activities is uneven. In the perspective of the mean $\mathcal{H}_{KS}$, the neural activities (including both spiking and resting) in shallow layers have intenser chaos, while those in deep layers have less intense chaos and are more stable. As verified by the normalized $\mathcal{H}_{KS}$, the stimulus-triggered responses of shallow-layer neurons are more regular and stable, while those of deep-layer neurons are more chaotic.

\begin{figure*}[!t]
\includegraphics[width=2\columnwidth]{GG-6.pdf}
\caption{Two findings about the dynamics effects on neural encoding and decoding. \textbf{(a)} The temporal and spatial distributions of $\mathcal{H}$, $\mathcal{H}^{\vartriangle}$, $\mathcal{H}^{\vartriangle\vartriangle}$, and $\mathcal{F}$. \textbf{(b)} The relations between $\Delta\mathcal{H}_{KS}$, $\Delta\mathcal{H}$, $\Delta\mathcal{H}^{\vartriangle}$, $\Delta\mathcal{H}^{\vartriangle\vartriangle}$, and $\Delta\mathcal{F}$. \textbf{(c)} The mean $\mathcal{H}$, $\mathcal{H}^{\vartriangle}$, $\mathcal{H}^{\vartriangle\vartriangle}$ and $\mathcal{F}$ (averaged through the time interval) are arranged based on the mean $\mathcal{H}_{KS}$. \textbf{(d)} Different from input neurons, the tuning curves of intermediary neurons (smoothed utilizing the Savitzky-Golay filter \cite{press1990savitzky} for visualization) usually feature more peaks (see left for an instance and see right for statistics). The stimuli located at the steep gradients around the peaks usually feature higher $\mathcal{F}$ (middle). Therefore, $\mathcal{F}$ relatively increases along with the number of peaks (right). \textbf{(e)} $x\in\lbrace10,20,30\rbrace$ bins (the length of each bin is $0.05$, $0.025$, or $0.0167$) are set for the mean $\mathcal{H}_{KS}$ interval. Then, the encoding ($\frac{\mathcal{H}^{\vartriangle\vartriangle}}{\mathcal{H}}$ and $\frac{\mathcal{H}^{\vartriangle\vartriangle}}{\mathcal{H}_{\mathcal{S}}}$) and decoding ($\mathcal{F}$) properties are averaged in bins. \textbf{(f)} The variation trends of the encoding scope and the local interpretability with the increasing mean $\mathcal{H}_{KS}$. Here we leave out an outlier data point that is away from the sample distribution.}
\end{figure*}
 
\subsection{Finding 3: The restrictive relationship between the encoding and decoding properties implied by dynamics}
 Now, we take information functions into analyses. Fig. 6(a) illustrates the temporal and spatial distributions of total entropy $\mathcal{H}$ (TE), noise entropy $\mathcal{H}^{\vartriangle}$ (NE), mutual information $\mathcal{H}^{\vartriangle\vartriangle}$ (MI) and Fisher information $\mathcal{F}$ (FI). In Fig. 6(b), we explore the relations between $\Delta\mathcal{H}_{KS}$ ($\tau=1$), $\Delta\mathcal{H}$, $\Delta\mathcal{H}^{\vartriangle}$, $\Delta\mathcal{H}^{\vartriangle\vartriangle}$, and $\Delta\mathcal{F}$, revealing that $\mathcal{H}$ and $\mathcal{H}^{\vartriangle}$ frequently share the same variation trends with $\mathcal{H}_{KS}$ (e.g., $\Delta\mathcal{H}\geq0$ and $\Delta\mathcal{H}^{\vartriangle}\geq0$ frequently hold when $\Delta\mathcal{H}_{KS}\geq 0$). This finding meets our expectation because $\mathcal{H}_{KS}$ and $\mathcal{H}$ both are the metrics of disorder and random degrees. Given the entropy of stimulus sequence $\mathcal{H}_{\mathcal{S}}$, the minimum noise entropy is bound by $\mathcal{H}^{\vartriangle}\geq\mathcal{H}-\min\left(\mathcal{H},\mathcal{H}_{\mathcal{S}}\right)$. Therefore, $\mathcal{H}^{\vartriangle}$ usually increases along with $\mathcal{H}$ when $\mathcal{H}_{\mathcal{S}}$ is given. However, the variation trends of $\Delta\mathcal{H}^{\vartriangle\vartriangle}$ and $\Delta\mathcal{F}$ can not be predicted by $\Delta\mathcal{H}_{KS}$ completely.
 
 To reveal the underlying patterns, we organize the experiment results as following: First, we average these parameters of each neuron through the time interval. Second, we arrange those averaged parameters of each neuron according to the mean $\mathcal{H}_{KS}$ (Fig. 6(c)).  Third, we do binning for $\mathcal{H}_{KS}$, and average $\mathcal{F}$, $\frac{\mathcal{H}^{\vartriangle\vartriangle}}{\mathcal{H}}$ (MI/TE) and $\frac{\mathcal{H}^{\vartriangle\vartriangle}}{\mathcal{H}_{\mathcal{S}}}$ (MI/ES) with respect to those bins (Fig. 6(d)). 

In Fig. 6(c), we can see that $\mathcal{H}$, $\mathcal{H}^{\vartriangle}$ and $\mathcal{H}^{\vartriangle\vartriangle}$ increase along with the mean $\mathcal{H}_{KS}$ (both for input and intermediary neurons), while $\mathcal{F}$ has more complex variation trends. The two-cluster distribution of $\mathcal{F}$ can be explained by the two-class neuron type (input and intermediary). This phenomenon is in line with our expectations. The mathematical definition of $\mathcal{F}$ (see (\ref{FisherInformation}) in appendix \ref{3.2M}) makes it depend on the intensity fluctuations of neural responses towards different stimuli. Consistent with previous studies \cite{dayan2001theoretical,butts2006tuning}, we discover that stimulus $s$ can cause large fluctuations when it is located at the steep gradients around the peaks of neural tuning curves, leading to a large $\mathcal{F}$ towards it (see Fig. 6(d)). Because the neural tuning curves of intermediary neurons usually feature more peaks, these neurons frequently have higher $\mathcal{F}$ than input neurons.

The results in Fig. 6(c-d) inspire us to further analyze the relationship between encoding and decoding properties. In Fig. 6(e), it can be seen that \textbf{(1)} $\mathcal{F}$, $\frac{\mathcal{H}^{\vartriangle\vartriangle}}{\mathcal{H}}$ and $\frac{\mathcal{H}^{\vartriangle\vartriangle}}{\mathcal{H}_{\mathcal{S}}}$ relatively increase when $\mathcal{H}_{KS}$ is in the range of $\left(0,0.2\right]$ (the left side of black dashed vertical line); \textbf{(2)} $\mathcal{F}$ and $\frac{\mathcal{H}^{\vartriangle\vartriangle}}{\mathcal{H}}$ decrease while $\frac{\mathcal{H}^{\vartriangle\vartriangle}}{\mathcal{H}_{\mathcal{S}}}$ continues increasing when $\mathcal{H}_{KS}$ is in $\left[0.2,0.5\right]$ (the right side of black dashed vertical line). Note that these phenomena do not depend on the binning approach critically. Overall, we conclude that when the dynamic randomness is relatively small ($\mathcal{H}_{KS}\in\left(0,0.2\right]$), the encoding (MI/ES) and decoding (FI) efficiency quantities share the same variation trend. When the dynamic randomness is relatively large ($\mathcal{H}_{KS}\in\left[0.2,0.5\right]$), an either-or situation emerges between encoding and decoding efficiency since these quantities have opposite trends along with $\mathcal{H}_{KS}$ (MI/ES increases but FI decreases). In other words, encoding efficiency has a restrictive relationship with the decoding efficiency under this condition since they can not be optimized synchronously.

\subsection{Finding 4: The relation between neural dynamics and the representation for stimulus distribution}
 We further analyze how neural dynamics determines the neural representation of the stimulus distribution. 

We have previously defined $\frac{\mathcal{H}^{\vartriangle\vartriangle}}{\mathcal{H}}$ to measure the interpretability of neural activities based on stimuli, or more specifically, based on the global stimulus distribution. Here the interpretability is defined in terms of the mutual information $\mathcal{H}^{\vartriangle\vartriangle}$, measuring the synergy degree between neural activities and stimuli. As illustrated in Fig. 4, neural activities become completely explainable when $\frac{\mathcal{H}^{\vartriangle\vartriangle}}{\mathcal{H}}$ approaches $1$.

To analyze the interpretability based on the local stimulus distribution, we introduce the conceptions of encoding scope $\mu$ and the local interpretability (Local MI/TE) (see (\ref{CSBases}-\ref{RecalculationDur}) in appendix \ref{3.1M}). Based on the definition of $\mathcal{H}^{\vartriangle}$, we can directly obtain the noise entropy $\mathcal{H}_{s}^{\vartriangle}$ that refers to each stimulus $s$. For $s$, the processing of it by neuron $N_{n}$ produces less noise if $\mathcal{H}_{s}^{\vartriangle}<\mathcal{H}^{\vartriangle}$, namely, it has better interpretability for the neural activities of $N_{n}$. Then, we define the encoding scope $\mu$ at moment $t$ as the proportion of the stimuli with better interpretability in all stimuli $\mathcal{S}^{\prime}\left(0,t\right)$
\begin{align}
\mu\left(N_{n},t\right)&=\dfrac{\vert\mathcal{S}_{\mu}\left(0,t\right)\vert}{\vert\mathcal{S}^{\prime}\left(0,t\right)\vert},\label{EQ9}\\
\mathcal{S}_{\mu}\left(0,t\right)&=\lbrace s\mid \mathcal{H}_{s}^{\vartriangle}\left(N_{n},t\right)<\mathcal{H}^{\vartriangle}\left(N_{n},t\right)\rbrace.\label{EQ10} 
\end{align}
When $\mu$ approaches $1$, it refers more to the global stimulus distribution; Otherwise, it refers more to specific local parts of stimulus distribution. By recalculating the total response entropy $\mathcal{H}_{\mu}$, noise entropy $\mathcal{H}_{\mu}^{\vartriangle}$, and mutual information $\mathcal{H}_{\mu}^{\vartriangle\vartriangle}$ only based on the stimuli in $\mathcal{S}_{\mu}$, we can further calculate the interpretability $\frac{\mathcal{H}_{\mu}^{\vartriangle\vartriangle}}{\mathcal{H}_{\mu}}$ of neural activities based on $\mathcal{S}_{\mu}$ (referred to as the local interpretability).

In Fig. 6(f), we show the variation trend of $\mu$ with respect to $\mathcal{H}_{KS}$, suggesting that $\mu$ is negatively correlated with $\mathcal{H}_{KS}$. Meanwhile, the local interpretability (Local MI/TE) relatively increases along with $\mathcal{H}_{KS}$. Taken together, we can conclude that when $\mathcal{H}_{KS}$ is in $\left(0,0.25\right]$ (the dynamics is more stable), neural activities can be better explained by the global stimulus distribution (larger encoding scope); Once the dynamic randomness becomes relatively large ($\mathcal{H}_{KS}$ is in $\left(0.25,0.5\right]$), neural activities can be better explained by the specific local parts of the stimulus distribution (smaller encoding scope). During this process, although the neurons with small dynamic randomness are mainly driven by the global stimulus distribution, they usually feature weaker neuron-stimulus synergy (lower Local MI/TE). This phenomenon is related to the low spiking probability of these neurons (see Fig. 5(b)), because their activation requires the inputs to contain enough global stimulus information (which can not be frequently satisfied). Opposite situations can be observed in the neurons with higher dynamic randomness, which feature stronger neuron-stimulus synergy (higher Local MI/TE).

 \begin{figure*}[!t]
\includegraphics[width=2\columnwidth]{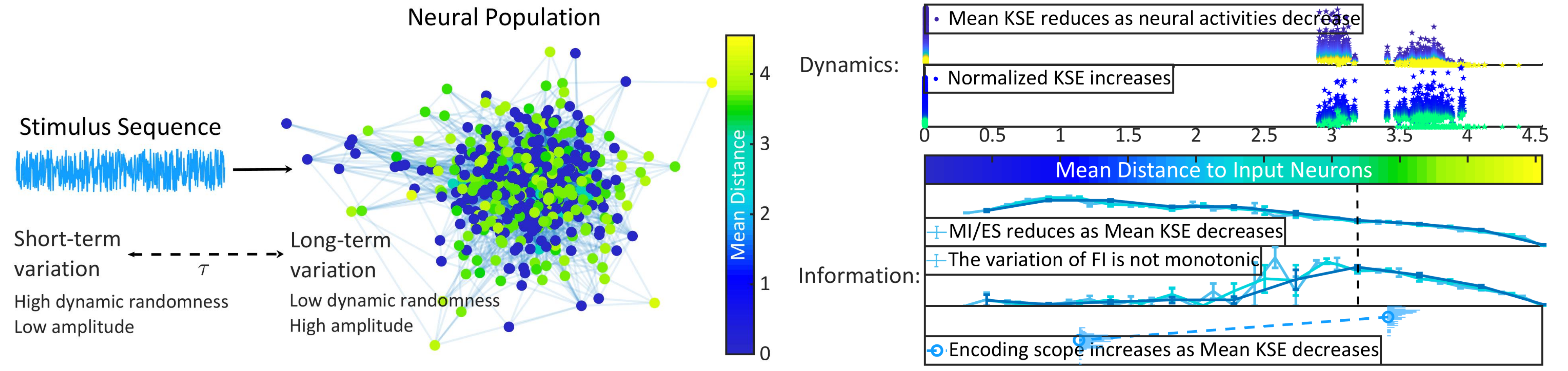}
\caption{Summary of our findings. A stimulus sequence triggers series of neural dynamics in a neural population. The dynamic variation of neural activities usually features small amplitudes and high dynamic randomness in the short-term, while the opposite properties can be seen in the long-term dynamic variation (see \textbf{Finding 1}). One can see an uneven spatial distribution of dynamic randomness in the neural population. Specifically, the mean $\mathcal{H}_{KS}$ declines from shallow-layer neurons to deep-layer neurons since neural spikes gradually reduce. Meanwhile, the normalized $\mathcal{H}_{KS}$ increases along with the mean distance to input neurons, suggesting that the chaotic degrees of stimulus-triggered neural responses increase from shallow-layer neurons to deep-layer neurons (see \textbf{Finding 2}). When one turns to analyze the information quantities, it can be found that the encoding efficiency $\frac{\mathcal{H}^{\vartriangle\vartriangle}}{\mathcal{H}_{\mathcal{S}}}$ (MI/ES) shares the same variation trend with the decoding efficiency $\mathcal{F}$ (FI) only when the dynamic randomness (mean $\mathcal{H}_{KS}$) is relatively small (in deep-layer neurons). When the dynamic randomness is relatively large (in shallow-layer neurons), the encoding efficiency increases along with the dynamic randomness while the decoding efficiency does not, leading to an either-or situation since the increase of the encoding/decoding efficiency implies the reduction of the other one (see \textbf{Finding 3}). Furthermore, the encoding scope gradually declines as the dynamic randomness increases. Thus, the shallow-layer neurons (with high dynamic randomness) mainly account for the specific encoding of local stimulus distribution, while the deep-layer neurons (with small dynamic randomness) support nonspecific encoding for the global stimulus distribution (see \textbf{Finding 4}). The observation of \textbf{Findings 2-4} requires the dynamic randomness to be analyzed in the short-term, otherwise the dynamic randomness will be small and homogeneous among all neurons (see \textbf{Finding 1}).} 
\end{figure*}

\section{Discussion}\label{Sec4}
\subsection{Significance of our work}
In the current research, we present an original theoretical framework and demonstrate the intrinsic connections between dynamics and information in neural activities. 

In the theoretical part, we propose the stimulus-triggered neural activity characterization as a bridge between the dynamics-theoretical and the information-theoretical metrics. We build our characterization only on several basic and common neural characteristics, such as tuning properties \cite{dayan2001theoretical,butts2006tuning} and neural spike mechanisms \cite{burkitt2006review}. These settings enable us to model real neural populations at a biologically authentic level. The proposed framework takes the advantages of both stochastic \cite{paninski2004maximum,heeger2000poisson,barbieri2001construction,burkitt2006review} and deterministic \cite{hodgkin1952quantitative,stein1965theoretical,izhikevich2003simple,gerstner2002spiking,tsodyks1998neural} models to offer a practical description of the collective neural activities involved in information processing. Specifically, we drive neural activities by the combined effects of stimulus-neuron synergy and network dynamics. While neural tuning properties directly govern the stimulus-neuron synergy, the network dynamics is captured by estimating the stochastic process of neurons from the neural activity samples generated by the neural response mechanism used in the deterministic models. Such an approach principally avoids the difficulty underlying a direct probabilistic description of the collective activities of coupled neurons. Based on this framework, the neural tuning Kolmogorov-Sinai entropy is introduced as a metric of the information-processing-related neural dynamics. Although Lyapunov exponents can not have classical definitions for the stochastic process since most trajectories in the phase space only spend a finite time in the system \cite{gaspard2004time}, a general connection between the entropy production and Lyapunov exponent can be established based on Ruelle inequality \cite{ruelle1978inequality} (or Pesin identity \cite{eckmann1985ergodic}), which enables the proposed neural tuning Kolmogorov-Sinai entropy to measure the dynamic randomness and identify chaos in the characterized neural activities analytically \cite{lecomte2007thermodynamic}. Taken together, the proposed framework supports calculating the dynamics-theoretical and the information-theoretical metrics analytically, achieving our objective for a unified analysis of dynamics and information. The analytical calculation of these metrics prevents our findings from depending on computational approximation critically.

In the experimental part, we implement a unified analysis for the relations between neural information functions (quantified from the aspects of encoding and decoding) and neural dynamics. We discover that short-term neural activities have smaller variation amplitudes but greater dynamic randomness, while long-term variations feature exactly opposite properties (\textbf{Finding 1}). Another relevant finding is that the spatial distribution of chaos of neural activities in a neural population is uneven (\textbf{Finding 2}). Then, we reveal the existence of specific restrictive relationships between the encoding and decoding efficiency when the dynamical randomness is relatively large (\textbf{Finding 3}). Finally, we identify that the neural activities with chaos dynamics are more related to the processing of local stimulus distribution while the stable neural dynamics is more relevant with the processing of global stimulus distribution (\textbf{Finding 4}). Fig. 7 offers a discussion on these findings and their relations.

\textbf{Finding 1} might be related to previous neural signal recording studies. It has been found that a signal recording scheme with a low temporal sampling rate (e.g., fMRI) can not directly reflect the underlying stimulus-triggered neural activities \cite{logothetis2001neurophysiological}. Before being recorded by those low temporal resolution techniques, the short-term neural activity variations need to accumulate until the variations are intense and robust enough \cite{logothetis2001neurophysiological}, accompanied by significant loss in neural activity information \cite{logothetis2001neurophysiological,lundengaard2017mechanistic,ekstrom2010and,singh2012neural}. A classical solution towards this limitation is to develop high temporal resolution recording techniques (e.g., multiphoton microscopy \cite{li2020adaptive,levene2004vivo,helmchen2002new}). While \textbf{Finding 1} partly supports this idea, it also suggests the contradiction between robust signal trends and the preservation capacity of the neural activity information. A recording scheme with a high sampling rate preserves the information of highly-frequent neural activity variations and ensures the separability of neurons, but obtains less robust signal trends than the scheme with a low sampling rate. Thus, improving temporal resolution is a necessary but not sufficient solution. 

If we do not control the effects of spiking probability, the variation trend of the mean $\mathcal{H}_{KS}$ in \textbf{Finding 2} is consistent with several findings in computational neuroscience. Back to an early study \cite{diesmann1999stable}, Diesmann finds an attractor of the propagation of synchronized action potentials, which governs the neural activity dynamics. A more recent study \cite{xiao2017cusps} confirms this attractor as a line attractor in the phase space of neural activities. These previous studies suggest that the dynamic randomness will be gradually reduced during a long enough propagation (e.g., from shallow layers to deep layers), accompanied by the reduction of spike rates. Compared with these previous results, our finding is not limited to the strictly hierarchical network topology and linear propagation process (our neural population is random), which ensures universality. When we turn to the normalized $\mathcal{H}_{KS}$ that quantifies the dynamic randomness of stimulus-triggered neural responses, one can see the consistency between its variation and a well-known phenomenon that stimulus drives suppress dynamic randomness \cite{molgedey1992suppressing}. Combine our results with these previous explorations \cite{molgedey1992suppressing}, we suggest that the synergy between neurons and input drives does control the chaos in stimulus-triggered neural responses (spikes). This neuron-stimulus synergy declines along with the distance to input neurons (the input ports of neural populations) and is gradually covered by network dynamics. The chaotic degree of neural responses to stimuli becomes larger when network dynamics takes in charge, because the stimulus effects are suppressed by inherent chaos. However, this phenomenon does not mean that the neural responses to stimuli are completely chaotic and unrepeatable (in other words, unreliable). As demonstrated by Fig. 6(d), the observed neural tuning curves of intermediary neurons are not fully stochastic, featuring specific patterns instead. Certain neural selectivity towards stimuli can still emerge. Therefore, the network dynamics and the inherent chaos can coexist with the regular neural activities governed by neural tuning properties. Although the regularity of neural activities is frequently broken by chaos, the chaotic degree of neural activities does not grow or maintain steadily (e.g., see Fig. 3(b)). These phenomena are consistent with the previous studies on chaos and reliability of stimulus-triggered neural activities \cite{lajoie2013chaos,lajoie2014structured}.

\textbf{Finding 3} may have potential insights for diverse topics, especially for the studies that aim at locating the neuronal or cortical foundations of cognitive functions \cite{passingham2002anatomical,gholipour2007brain,purves2008cognitive}. These studies analyze the information processing properties of specific neurons or brain regions. The analysis usually relies on specific signal recording schemes (e.g., multiphoton microscopy \cite{reddy2008three,dunn2008functional} and microelectrode recording \cite{ulbert2001multiple,cox2008high}) and measures the information processing properties from the aspects of encoding (e.g., with $\mathcal{H}$, $\mathcal{H}^{\vartriangle}$, $\mathcal{H}^{\vartriangle\vartriangle}$ \cite{borst1999information,nemenman2008neural}) or decoding (e.g., with Fisher information $\mathcal{F}$ \cite{wu2002population,quiroga2009extracting,toyoizumi2006fisher}). Based on \textbf{Finding 3}, the potential risks lie in that the direct and indirect measurement methods may obtain two separate, and even competing, results. For example, on the one hand, the real efficient neurons/cortices in the encoding process might be neglected since its recorded signals are measured with low decoding efficiency; on the other hand, the neurons/cortices that are efficient in decoding might have lots of activities that can not be explained by stimuli, and thus, imply noisy conditions in the analysis. These risks may further lead to false-negative problems and repeatability problems in functional studies \cite{brett2002problem}.

Moreover, \textbf{Finding 1} and \textbf{Finding 3} might propose challenges for neural signal recordings by revealing \textbf{(1)} the contradictory relation between the robust signal trends and the preservation capacity of the neural activity information in the recording stage and \textbf{(2)} the restrictive relations between the encoding and decoding properties in the analysis of recorded signals. Although the recording with high temporal resolution features the capacity to reflect the underlying frequent neural activity variations, the high dynamic randomness of short-term neural activities usually limits the possibility to obtain a robust and repeatable result and implies the restrictive relations between encoding and decoding properties. The recording scheme with low temporal resolution can not record the high-frequency neural activities, but it can obtain the signals with relatively (not completely) controlled dynamic randomness to ensure stable macroscopic signal trends and avoid the restrictive relations between encoding and decoding. Therefore, a multi-temporal-resolution recording may take the advantages of both high and low temporal resolutions and overcome their shortcomings. This provides theoretical interpretations for the intrinsic advantages of the multi-temporal-resolution recording of neural signals \cite{qin2004multi,li2011time,indiradevi2008multi}. 

\textbf{Finding 4} might be inspiring for cognitive neuroscience. Both \textbf{Finding 2} and \textbf{Finding 4} reveal that the neural activities in shallow layers are more related to local stimulus distribution, while those in deep layers are more related to the global stimulus distribution. To some extent, the interpretability by the global stimulus distribution can be treated as a kind of increase of the generalization ability of neural responses to stimuli. This emerged phenomenon suggests the possibility that the neurons in shallow layers concentrate on specific local information of stimuli, while the neurons in deep layers process the general information of stimuli. To our knowledge, there is no experimental evidence for such a finding, and it may inspire further study on the spontaneous formation of the division of labor among neurons during the data-driven (bottom-up) processing.

\subsection{Validity and limitations}
In the above discussion, we have sketched how our findings may relate to neuroscience studies. Here we attempt to confirm the scope that our work can be applied on robustly and validly. Although we have pursued a unified and biologically valid analysis of information and dynamics, there are specific limitations in the current research. The validity of the mathematical descriptions of neural characteristics (e.g., tuning properties \cite{dayan2001theoretical,butts2006tuning} and neural spike mechanisms \cite{burkitt2006review}) is ensured conditionally.

The necessary condition for the proposed Poisson processes to approximate neural activities validly is that the time step $\Delta t$ in discretization corresponds to a sufficiently short physical time ($\sim 5$ ms). With a sufficiently small physical time step, the activities of every neuron follow a non-homogeneous Poisson process because a neuron can not emit more than one spike simultaneously (satisfies the Poisson condition that the probability of two or more changes in a sufficiently small interval is $0$). Although the dependence on short physical time steps does not threaten our theory and findings mathematically, it is essentially a limitation in computational implementations. The current version of stimulus-triggered neural activity characterization is extremely computationally costly, making it impossible to deploy large-scale and long-run experiments (e.g., run on an ultra-large neural population with $10^9$ neurons that approximates a human cortical area or generate the spike trains corresponding to a physical time interval of $24$ hour). The costs will significantly increase if we further take the maximum likelihood estimation for defining the Poisson processes of intermediary neurons (see (\ref{InterveingIntensityF})) into account. However, the formation of neural characteristics (e.g., the tuning properties of intermediary neurons) in real neural systems usually requires long-term processes and the involvements of large-scale neurons. The difficulty we meet here is a potential threat to the generalization ability of our findings, questioning if the discovered phenomena rely on our experiment settings critically.

Proposing mechanistic insights into why our findings arise is a possible approach to verify their generalization ability. \textbf{Finding 1} arises when we attempt to compare between the short-term (small $\tau$) and long-term (large $\tau$) variations of neural activities. As suggested by Fig. 3(a) and Fig. 5(a), the differences between short-term and long-term neural activity variations (amplitude and dynamic randomness) hold across different neurons and throughout the time interval. The amplitude differences are easy to understand since the variation amplitude accumulates during the variation interval $\left[t,t+\tau\right]$. A larger $\tau$ naturally implies larger accumulations of amplitudes. The difference relates to dynamic randomness mainly arises from the mathematical definition of neural tuning Kolmogorov-Sinai entropy $\mathcal{H}_{KS}$ (see (\ref{NKSE})). One can reorganize (\ref{NKSE}) as 
\begin{align}
&\mathcal{H}_{KS}\left(N_{n},\mathcal{S}^{\prime},t,t+\tau\right)=\sum_{r}\mathcal{P}_{n}^{\heartsuit}\left(r\mid\mathcal{S}^{\prime},0,t\right)\notag\\
&\times\left(\sum_{r^{\prime}> r}\mathcal{W}_{rr^{\prime}}\left(\mathcal{S}^{\prime},t,t+\tau\right)\ln\sqrt[\tau]{\frac{1}{\mathcal{W}_{rr^{\prime}}\left(\mathcal{S}^{\prime},t,t+\tau\right)}}\right),\label{EQ11}
\end{align}
where the second part relates to $\tau$ and is subject to $\sum_{r^{\prime}\geq r}\mathcal{W}_{rr^{\prime}}\left(\mathcal{S}^{\prime},t,t+\tau\right)=1$. Based on information theory, the term $\sum_{r^{\prime}> r}\mathcal{W}_{rr^{\prime}}\left(\mathcal{S}^{\prime},t,t+\tau\right)\ln\frac{1}{\mathcal{W}_{rr^{\prime}}\left(\mathcal{S}^{\prime},t,t+\tau\right)}$ will be maximized if the probability distribution $\lbrace\mathcal{W}_{rr^{\prime}}\rbrace_{r^{\prime}}$ approaches the uniform distribution on $\left[r,\widehat{r}\right]$ (here $\widehat{r}$ denotes the maximum response rate that can be reached by the neuron). This is because the uniform distribution is the maximum entropy distribution defined on finite interval and has no constraint on moments \cite{cover1999elements}. As shown by Fig. 3(a), the distribution $\lbrace\mathcal{W}_{rr^{\prime}}\rbrace_{r^{\prime}}$ tends to approach the uniform distribution as $\tau$ increases (distribution peaks become broader). Therefore, one can verify that $\sum_{r^{\prime}> r}\mathcal{W}_{rr^{\prime}}\left(\mathcal{S}^{\prime},t,t+\tau\right)\ln\frac{1}{\mathcal{W}_{rr^{\prime}}\left(\mathcal{S}^{\prime},t,t+\tau\right)}$ increases along with $\tau$. However, the actual second part in (\ref{EQ11}) is $\sum_{r^{\prime}> r}\mathcal{W}_{rr^{\prime}}\left(\mathcal{S}^{\prime},t,t+\tau\right)\ln\sqrt[\tau]{\frac{1}{\mathcal{W}_{rr^{\prime}}\left(\mathcal{S}^{\prime},t,t+\tau\right)}}$, featuring an opposite variation trend. The increase will be reversed by the $\tau$-th root $\sqrt[\tau]{}$ and becomes decrease, implying that entropy $\mathcal{H}_{KS}$ reduces along with $\tau$. In summary, \textbf{Finding 1} mainly emerges from the mathematical nature of the proposed $\mathcal{H}_{KS}$. 

\textbf{Finding 2} arises when we attempt to compare the dynamic randomness between neurons. In Fig. 5(b), the mean $\mathcal{H}_{KS}$ declines along with the mean distance to input neurons because neural spike rates reduce. Similar with \textbf{Finding 1}, this phenomenon results from the mathematical definition of entropy $\mathcal{H}_{KS}$ as well. One can verify that $\mathcal{H}_{KS}$ increases along with $\mathcal{P}_{n}^{\heartsuit}\left(r\mid\mathcal{S}^{\prime},0,t\right)$ (see (\ref{EQ11})), while the latter governs the spiking probability of neurons. Thus, the dynamic randomness will increase if the neuron tends to emit spikes. In our experiment, we also calculate the normalized $\mathcal{H}_{KS}$ (averaged from the raw data of $\mathcal{H}_{KS}$ corresponding to neural spikes and normalized by spiking probability). By controlling the effects of spiking probability, the observed variation trend of the normalized $\mathcal{H}_{KS}$ is consistent with previous studies that stimulus drives suppress dynamic randomness (or chaotic degrees) \cite{molgedey1992suppressing}. These phenomena can be related to the mathematical properties of $\mathcal{H}_{KS}$ in general.

\textbf{Finding 3} arises when we analyze the encoding and decoding properties as the functions of dynamic randomness. For encoding properties, one can see an increasing encoding efficiency (MI/TE) along with the dynamic randomness (measured by the mean $\mathcal{H}_{KS}$). Although this phenomenon is consistent with the common belief that representing variable stimulus information requires high variability of neural activities, it can not be derived from the mathematical definition of mutual information directly. For decoding properties, we have demonstrated that input neurons (with higher dynamic randomness) can feature less Fisher information (FI) than intermediary neurons (with lower dynamic randomness), which mainly results from the differences between the tuning properties of these two kinds of neurons (see Fig. 6(d)). Therefore, the decoding efficiency does not necessarily increase along with the mean $\mathcal{H}_{KS}$. Limited by the size of our experiments, the observed phenomenon in Fig. 6(e) is that the decoding efficiency (FI) increases when the mean $\mathcal{H}_{KS}$ is relatively small and decreases when the randomness is sufficiently large. If the analysis is implemented on a sufficiently large neuron population (the variation range of the mean $\mathcal{H}_{KS}$ is enlarged) that processes stimuli in a sufficiently long interval (the tuning curves of intermediary neurons become more smooth), we hypothesize that the variation trend of decoding efficiency will become more smoother and completely decreasing along with the mean $\mathcal{H}_{KS}$. Meanwhile, the quantity gap between input neurons and intermediary neurons in decoding efficiency will decrease to a reasonable range. In brief, \textbf{Finding 3} mainly arises from neural tuning properties rather than the mathematical attributes of the proposed information-theoretical metrics. Although we hypothesize that \textbf{Finding 3} principally holds in most cases, any generalization of \textbf{Finding 3} should be verified carefully. 

\textbf{Finding 4} arises when we explore the neural representation of the stimulus distribution. In our research, we propose a new conception, the encoding scope $\mu$, to capture the characteristics of neural representation. The encoding scope $\mu$ principally measures the proportion of stimulus sub-set $\mathcal{S}_{\mu}$ that has better interpretability for neural activities (the noise entropy quantities of encoding them are below average) in all stimuli. In Fig. 6(f), we have observed that the encoding scope decreases along with the mean $\mathcal{H}_{KS}$, implying that shallow-layer neurons have smaller encoding scope than deep-layer neurons. Moreover, the local interpretability (the interpretability of neural activities by the stimuli within $\mathcal{S}_{\mu}$) decreases from shallow-layer neurons to deep-layer neurons. In general, we suggest that these phenomena result from the mathematical nature of noise entropy (see (\ref{NE}) and (\ref{NEofS})) as well as the differences between shallow-layer and deep-layer neurons in neural tuning properties. Please note that the noise entropy towards stimulus $s$ (see (\ref{NEofS})) is defined as $\mathcal{H}_{s}^{\vartriangle}\left(N_{n},t\right)=-\sum_{r}\mathcal{P}_{n}^{\heartsuit}\big(r\mid s,0,\tau_{\mathcal{S}}\left(t\right)\big)\log_{2}\mathcal{P}_{n}^{\heartsuit}\big(r\mid s,0,\tau_{\mathcal{S}}\left(t\right)\big)$. Following a similar idea that we have applied on \textbf{Finding 1}, the entropy quantity will be maximized if the probability distribution $\lbrace\mathcal{P}_{n}^{\heartsuit}\big(r\mid s,0,\tau_{\mathcal{S}}\left(t\right)\big)\rbrace_{r}$ approaches the uniform distribution on $\left[0,\widehat{r}\right]$ (e.g., the peaks of distribution becomes broader). Because $\mathcal{P}_{n}^{\heartsuit}\big(r\mid s,0,\tau_{\mathcal{S}}\left(t\right)\big)$ is governed by the tuning curve $\mathcal{G}_{n}$ of neuron $N_{n}$, the approaching process essentially requires the response coefficient $\mathcal{G}_{n}\left(s\right)$ to be large. Otherwise the density of $\lbrace\mathcal{P}_{n}^{\heartsuit}\big(r\mid s,0,\tau_{\mathcal{S}}\left(t\right)\big)\rbrace_{r}$ will concentrate on a narrow sub-interval of $\left[0,\widehat{r}\right]$ where $r$ is small. One can verify that shallow-layer neurons (e.g., input neurons) usually feature broader tuning curve peaks and higher maximum response coefficients than deep-layer neurons (e.g., see Fig. 6(d)). For shallow-layer neurons, this property makes the noise entropy towards a wider range of stimuli located near tuning curve peaks sufficiently large, leading to higher noise entropy quantities (see Fig. 6(c)). Meanwhile, there remain relatively few stimuli with low response coefficients, implying a small size of $\mathcal{S}_{\mu}$ because most stimuli correspond to high noise entropy. Therefore, the encoding scopes of shallow-layer neurons are frequently small. Opposite situations can be seen in deep-layer neurons because of their relatively low response coefficients to stimuli and narrower tuning curve peaks. As for the differences between shallow-layer and deep-layer neurons in local interpretability, although we hypothesize that this phenomenon arises from neural tuning properties, we can not derive it mathematically in the current work.

In summary, we suggest that \textbf{Finding 1} and \textbf{Finding 2} emerge from the mathematical properties of the neural tuning Kolmogorov-Sinai entropy $\mathcal{H}_{KS}$ and keep consistency with neural characteristics. The validity and generalization ability of these two findings can be partly ensured. As for \textbf{Finding 3} that results from neural tuning properties, we hypothesize that it principally holds under different conditions because the involved neural tuning properties are basic properties of neural systems. However, we need to emphasize that our results have not been verified in large-scale and long-run experiments. The phenomenon itself, as well as related discussions, should be treated carefully. For \textbf{Finding 4}, although the phenomenon relevant with encoding scope can be mathematically derived from noise entropy and neural tuning properties, the phenomenon related to local interpretability remains a subject for further investigation.

\subsection{Future directions}
Understanding the connection between dynamics and information in the brain has become one of the most critical challenges in physics and neuroscience, leading to a promising way to improve the interpretability of information and cognitive functions based on physical foundations \cite{vogels2005neural}. 

Our theoretical framework is a possible paradigm towards the unified measurement and analysis of dynamics and information attributes of neural activities. As suggested above, more verification is necessary for the potential connections between information and dynamics identified by this framework. In future works, one can further explore our framework in the aspect of the stochastic dynamics of the master equation \cite{sekimoto2010stochastic,seifert2012stochastic} (or known as the Schnakenberg network theory \cite{schnakenberg1976network,andrieux2007fluctuation}). Another valuable direction in the theoretical analysis is to further study the potential connections between the proposed neural tuning Kolmogorov-Sinai entropy and the Lyapunov spectra (one can turn to \cite{engelken2020lyapunov,sompolinsky1988chaos} for the applications of the Lyapunov spectra in neural network studies). Built on the current qualitative connection established by Ruelle inequality \cite{ruelle1978inequality} (or Pesin identity \cite{eckmann1985ergodic}), more intrinsic properties of neural dynamics might be revealed if a systematic unification is implemented between these two kinds of metrics in neural activities at a more quantitative level. Moreover, although the roles of neural plasticity (e.g., STDP \cite{caporale2008spike,dan2004spike}) have not been included in our current analysis, the proposed stimulus-triggered neural activity characterization can be easily generalized to plasticity conditions to study the effects of memory (e.g., defining the neural response threshold $\Upsilon$ in (\ref{Threshold}) as time-dependent). It is expected that this unified framework features the potential to deepen our understanding of the emergence of various characteristics of neural information processing from physical bases.

In the current research, we have focused on implementing the unification on the sub-cortex scale (e.g., a neural population with thousands of neurons), where each neuron plays a critical role and the cognitive functions have not emerged yet. Our future pursuit is to generalize the present framework to the cortex scale and further analyze cognitive functions. On the cortex surface, the ultra-dense distributions of neurons and synapses make the roles of the individual neuron or local neural network topology be covered up by the global cortex dynamics during the information processing process \cite{gerstner2014neuronal}. Therefore, the challenge we face is to develop a macroscopic description of the information-processing-related neural dynamics. One possible approach is to implement the renormalization group \cite{fisher1974renormalization,weinberg1973new,koch2018mutual} on neural dynamics and analyze the multi-scale dynamics transformation. Such analysis allows us to understand the accumulation of dynamics from cellular scale to cortical scale. Another potential way is to develop a continuous formulation of neural dynamics by approaching the thermodynamic limit of the proposed neural activity characterization in our research \cite{montbrio2015macroscopic}. Then, we can combine our characterization framework with specific cortical field models of perception (e.g., the ring models of primary visual cortex \cite{ben1995theory,shriki2003rate}) to generate the information-processing-related dynamics in the ultra-large neural population of certain sensory cortices. Based on these implementations, it might be possible for our framework to be applied in further studying how cognitive functions are shaped by global cortex dynamics. 

In the viewpoint of Charles F. Stevens \cite{koch2000role},  models are common in neuroscience, but theories are relatively scarce. Neuroscience has amassed various models to describe specific phenomena, but few theories offer general frameworks for a wide range of facts and find the underlying connections between different issues. Our research, as well as diverse previous explorations (e.g., see the works on neural information functions \cite{averbeck2006neural} and the works on neural dynamics \cite{chialvo2010emergent}), demonstrate a possible way to identify the general connections between information and dynamics in the brain. These present theories and discoveries suggest an evolutionary perspective that the characteristics of neural information functions and further cognitive functions naturally emerge from the physically fundamental properties of neural ensembles rather than be designed by complex high-level mechanisms. This suggested perspective is worthy of further explorations in the future. 

\section{Acknowledgements}
The authors are grateful to Ms. Connie Marker for her help in manuscript preparation. This project is supported by the Artificial and General Intelligence Research Program of Guo Qiang Research Institute at Tsinghua University (2020GQG1017) as well as the Tsinghua University Initiative Scientific Research Program. 

\begin{appendix}
 \section{The stochastic process of neural activities}
\subsection{Characterize neural populations}\label{1.1M}

We begin with a random neuron population $\mathcal{N}\left(\mathcal{V},\mathcal{E}\right)$, where $\mathcal{V}$ is the set of all neurons and $\mathcal{E}$ is the set of all synapses. We use the weighted adjacent matrix $\mathcal{C}$ to describe the synaptic connection strength between any two neurons $N_{i}$ and $N_{j}$ as $\mathcal{C}_{ij}\in\left[-1,1\right]$ (here $\mathcal{C}_{ij}<0$ stands for the inhibitory connection, and $\mathcal{C}_{ij}>0$ stands for the excitatory connection, and $\mathcal{C}_{ij}=0$ means that there is no synaptic connection). In experiments, we randomly generate $\mathcal{V}$, $\mathcal{E}$, and $\mathcal{C}$ for university. Specifically, the randomization of $\mathcal{E}$ utilizes a basic approach introduced by Erd{\H{o}}s and R{\'e}nyi \cite{erdHos1960evolution,cameron1997random}. The probability for any two neurons to feature a synaptic connection is set as $p$, implying that the average degree of neurons equals $p\left(\vert\mathcal{V}\vert-1\right)$. For convenience, our research randomize $p\in\left[0.02,0.025\right]$ in every experiment. As for $\mathcal{C}$, each element in it is uniformly randomized from $\left[-1,1\right]$.

In a neural population, the stimulus inputs can not be simultaneously received by all the neurons. We refer to the neurons that receive inputs directly and instantaneously as the input neurons. As for the neurons that are not directly triggered by stimulus inputs, they can be activated by the stimulus information transmitted from the neurons in its receptive field (the set of its pre-synaptic neurons). We call them intermediary neurons. To keep our experiments universal, we randomly pick a subset of neurons in a generated neural population as input neurons. As for the remaining neurons, they are considered as intermediary neurons. 
 
\subsection{The neural activity of input neuron}\label{1.2M}
Each input neuron $N_{i}$ processes stimuli directly, whose activity profile is appropriately shaped by its tuning properties (i.e., response selectivity to stimuli). Neural tuning curve is a standard model to describe the response selectivity \cite{dayan2001theoretical,butts2006tuning}, where the stimulus at the peak evokes the highest response rate \cite{butts2006tuning}. A representative example is the bell-shaped tuning curve \cite{dayan2001theoretical,butts2006tuning}, which is
\begin{equation}
\mathcal{G}_{i}\left(s\right)=\widehat{R}_{i}\exp\left(-0.5\left(\frac{s-\widehat{s}_{i}}{\sigma_{i}}\right)^2\right), \label{TuningCurve}
\end{equation}
where $\widehat{R}_{i}$ is the maximum response coefficient, $\widehat{s}_{i}$ is the preferred stimulus selected from the stimulus set $\mathcal{S}$, and $\sigma_{i}$ represents the width of the tuning curve. 

Assume that a stimulus sequence $\mathcal{S}^{\prime}$ occurs in a given time interval $\left[0,t^{\prime\prime}\right]$. For convenience, sequence $\mathcal{S}^{\prime}$ is sampled from $\mathcal{S}$ uniformly in our experiments, and the time unit $\Delta t$ (minimum time step for discretization) is set as $1$. Note that other kinds of randomization and discretization can also be applied.

Following the perspective of rate coding \cite{dayan2001theoretical}, we implement the neural activity characterization as the probability for the neural response to reach a specific frequency at a given moment. The Poisson process is efficient in approximating this probability in most cases \cite{dayan2001theoretical}. Given the stimulus sequence $\mathcal{S}^{\prime}$, the probability for the cumulative neural response count of an input neuron $N_{i}$ to reach a specific quantity $r$ at moment $t$ ($t\in\left[0,t^{\prime\prime}\right]$) can be approximated by a non-homogeneous Poisson process, whose probability distribution is
\begin{align}
\mathcal{P}_{i}\left(r\mid \mathcal{S}^{\prime},0,t\right)=\dfrac{\left[\Lambda_{i}\left(0,t\right)\right]^{r}}{r!}\exp\left(-\Lambda_{i}\left(0,t\right)\right),\label{MasterEq}
\end{align}
where $\Lambda_{i}\left(0,t\right)$ is the cumulative intensity function
\begin{equation}
\Lambda_{i}\left(0,t\right)=\int_{0}^{t}\lambda_{i}\left(m\right)dm, \label{CIF}
\end{equation}
and $\lambda_{i}\left(m\right)$ denotes the time-varying intensity function
\begin{equation}
\lambda_{i}\left(m\right)=\mathcal{G}_{i}\left(\mathcal{S}^{\prime}\left(m\right)\right). \label{TDF}
\end{equation}
Given (\ref{MasterEq}-\ref{TDF}), the activity profile of input neuron $N_{i}$ is defined based on the interactions between the stimulus sequence $\mathcal{S}^{\prime}$ and the neural tuning properties $\mathcal{G}_{i}\left(s\right)$.

Given the above definition, we can further generate possible neural activities. This research generates neural activities by predicting the arrival time sequence of neural responses with the maximum probability method (MP). For the $r$-th neural response, we can find the location of the maximum probability of it by working out 
\begin{equation}
\widehat{t}_{r}=\operatorname*{argmax}_{t}\left(\mathcal{P}_{i}\left(r\mid \mathcal{S}^{\prime},0,t\right)\right). \label{MPM}
\end{equation}
The obtained result $\widehat{t}_{r}$ by the operator $\operatorname*{argmax}$ is the moment that maximizes the probability $\mathcal{P}_{i}\left(r\mid \mathcal{S}^{\prime},0,t\right)$, meaning that the probability for the $r$-th neural response to arrive at moment $\widehat{t}_{r}$ is highest. Therefore, the $r$-th neural response high-frequently occurs at moment $\widehat{t}_{r}$ in real situations. The MP method benefits neural response generation for its low computational costs and its ability to approximate the neural response generation by large-scale Monte Carlo sampling \cite{shapiro2003monte} on probability distributions (maximum probability implies the highest occurrence frequency when the sampling size is large enough). One can replace the MP method with large-scale Monte Carlo sampling when computing power allows. A potential limitation of the MP method lies in that it essentially creates a deterministic mapping from the probability distribution $\mathcal{P}_{i}\left(r\mid \mathcal{S}^{\prime},0,t\right)$ to the neural response. Although this property does not affect our research because neural response trains that strictly reflect $\mathcal{P}_{i}\left(r\mid \mathcal{S}^{\prime},0,t\right)$ are exactly demanded in our analysis, the MP method is not applicable when one demands more randomness in neural response generation (e.g., to simulate noisy neural responses that do not follow $\mathcal{P}_{i}\left(r\mid \mathcal{S}^{\prime},0,t\right)$ strictly). In the latter situation, a relatively small-scale Monte Carlo sampling can be used to create more randomness.

By traversing all possible response rate $r$, we can obtain a set of moments $\lbrace \widehat{t}_{r}\rbrace_{r\in\mathbb{N}^{+}}$ based on (\ref{MPM}). Given the properties of Poisson process, we know that $\lbrace \widehat{t}_{r}\rbrace_{r\in\mathbb{N}^{+}}$ is naturally ensured to be not decreasing. For convenience, we mark every neural response in the timeline following
\begin{equation}
\mathcal{R}_{i}\left(t\right)=\sum_{\widehat{t}_{r}\in\mathcal{T}_{i}}\delta\left(t-\widehat{t}_{r}\right), \label{MarkerofNR}
\end{equation}
where $\mathcal{T}_{i}=\lbrace\widehat{t}_{r}\rbrace_{r\in\mathbb{N}^{+}}$ and $\delta$ denotes Dirac Delta function. The obtain sequence $\mathcal{R}_{i}\left(t\right)$ equals $+\infty$ if a neural response arrives at moment $t$. Otherwise it equals $0$. Please note that one should replace all $+\infty$ in $\mathcal{R}_{i}$ by $1$ in computational implementations. 

\subsection{The neural activity of intermediary neuron} \label{M1.3}
Each intermediary neuron $N_{j}$ is driven by the pre-synaptic neurons $\mathcal{RF}\left(N_{j}\right)$ and affected by the network dynamics, whose activities can not be simplified as (\ref{MasterEq}). This research characterizes the neural activity profile of $N_{j}$ by describing its synaptic inputs and neural responses to these inputs. 

For each neuron $N_{k}$ in $\mathcal{RF}\left(N_{j}\right)$, let its neural response arrival sequence be $\widehat{\mathcal{R}}_{k}\left(t\right)$, then we can describe the synaptic input given from $N_{k}$ to $N_{j}$ as $\widehat{\mathcal{R}}_{k}\left(t\right)\mathcal{C}_{kj}$. In most cases, the cumulative synaptic inputs on neuron $N_{j}$ have a dissipation term $\Omega_{j}\left(t\right)$ (e.g., the leaky term in the leaky integrate-and-fire neuron \cite{burkitt2006review}). And $N_{j}$ emits a neural response only if the cumulative synaptic inputs reach a specific response threshold $\Upsilon_{j}\left(t\right)$ (e.g., spiking threshold \cite{burkitt2006review}). Therefore, we can define the neural response of $N_{j}$ as
\begin{align}
\widehat{\mathcal{R}}_{j}\left(t\right)=\upsilon\left[\int_{0}^{t}\Big(\Psi_{j}\left(x\right)+\Omega_{j}\left(x\right)\Big)dx-\Upsilon_{j}\left(t\right)\right], \label{PostNeuralResponse}
\end{align}
where $\upsilon\left(\cdot\right)$ denotes the unit step function, and $\Psi_{j}\left(x\right):=\sum_{N_{k}\in\mathcal{RF}\left(N_{j}\right)}\widehat{\mathcal{R}}_{k}\left(x\right)\mathcal{C}_{kj}$ denotes the synaptic inputs. In our experiments, the definitions of $\Omega_{j}\left(t\right)$ and $\Upsilon_{j}\left(t\right)$ are proposed in general forms. Specifically, $\Omega_{j}\left(t\right)$ satisfies
\begin{align}
\int_{0}^{t}\Omega_{j}\left(x\right)dx=-\upsilon\left(t-\widehat{\tau}\right)\int_{0}^{t-\widehat{\tau}}\Psi_{j}\left(x\right)dx+\varepsilon\left(t\right),\label{DT}
\end{align}
where $\widehat{\tau}$ denotes the minimum time consumption of generating spikes and $\varepsilon\left(t\right)$ is the time-dependent perturbation. Based on (\ref{DT}), we can realize that \textbf{(1)} for $t\in\left[0,\widehat{\tau}\right)$, the cumulative synaptic inputs of $N_{j}$ during $\left[0,t\right]$ will only be perturbed by $\varepsilon\left(t\right)$ rather than dissipated. This is natural because the $1$-st spike generation does not end yet and historical accumulations should not dissipate; \textbf{(2)} for $t\in\left[\widehat{\tau},t^{\prime\prime}\right]$, the cumulative synaptic inputs of $N_{j}$ during $\left[0,t\right]$ will be dissipated to $\int_{t-\widehat{\tau}}^{t}\Psi_{j}\left(x\right)dx+\varepsilon\left(t\right)$ at moment $t$ by the dissipation term. This definition ensures that the $k$-th spike generation ($k>1$) is driven by the cumulative synaptic inputs during $\left[t-\widehat{\tau},t\right]$ and a small quantity of historical perturbations remaining for dissipation. In computational implementations, we can do discretization (set the minimum time step $\Delta t=\widehat{\tau}$) on (\ref{DT}) to realize that historical accumulations before the $\left(n-1\right)$-th step will dissipate at the $n$-th step and only leave behind specific perturbations. A concrete example of the general definition in (\ref{DT}) is the leaky term in the leaky integrate-and-fire neuron \cite{burkitt2006review}, where the time-dependent perturbation $\varepsilon\left(t\right)$ is frequently omitted. In our research, this perturbation is defined as
\begin{align}
\varepsilon\left(t\right)\in\left[-\frac{1}{\gamma_{\varepsilon}}\max_{t}\Psi_{j}\left(t\right),\;\frac{1}{\gamma_{\varepsilon}}\max_{t}\Psi_{j}\left(t\right)\right], \label{Perturbation}
\end{align}
in which $\gamma_{\varepsilon}$ is the degree of perturbation. In our experiment, we randomly define $\gamma_{\varepsilon}\in\left[20,50\right]$ to control the intensity of perturbation. As for the time dependent neural response threshold $\Upsilon_{j}\left(t\right)$, it can be defined in various forms to achieve plasticity mechanisms \cite{caporale2008spike,dan2004spike}. In our experiment, we use a simplified definition
\begin{equation}
\forall t,\;\Upsilon_{j}\left(t\right)=\theta\max_{t}\Psi_{j}\left(t\right), \label{Threshold}
\end{equation}
in which we randomly set $\theta\in\left[\frac{1}{4},\frac{3}{4}\right]$ based on the ratio of the difference between the response threshold and the resting state to the difference between the response apex and the resting state \cite{yi2015biophysical,izhikevich2003simple,izhikevich2004model}. 

Based on (\ref{PostNeuralResponse}), we can obtain an observed neural response sequence of the intermediary neuron $N_{j}$. We repeat the experiment with the stimulus sequence $\mathcal{S}^{\prime}$ for $h$ times, each time we can obtain an observed result $\widehat{\mathcal{R}}_{j,a}\left(t\right)$ ($a\in \mathbb{Z}\cap\left[1,h\right]$). Then we can use the maximum likelihood estimation to construct the observed non-homogeneous Poisson process of the neural activities of $N_{j}$, where the observed time-varying intensity function $\widehat{\lambda}_{j}\left(t\right)$ is given as
\begin{equation}
\widehat{\lambda}_{j}\left(t\right)=\frac{1}{h}\sum_{a\in \mathbb{Z}\cap\left[1,h\right]}\widehat{\mathcal{R}}_{j,a}\left(t\right). \label{InterveingIntensityF}
\end{equation}
Based on (\ref{InterveingIntensityF}), the observed Poisson process of $N_{j}$ features a probability distribution
\begin{align}
\widehat{\mathcal{P}}_{j}\left(r\mid \mathcal{S}^{\prime},0,t\right)=\dfrac{\left[\widehat{\Lambda}_{j}\left(0,t\right)\right]^{r}}{r!}\exp\left(-\widehat{\Lambda}_{j}\left(0,t\right)\right),\label{MasterEqOb}
\end{align}
where $\widehat{\Lambda}_{j}\left(0,t\right)$ is the observed cumulative intensity function, 
\begin{equation}
\widehat{\Lambda}_{j}\left(0,t\right)=\int_{0}^{t}\widehat{\lambda}_{j}\left(m\right)dm. \label{ObIF}
\end{equation}
Given (\ref{MasterEqOb}), the MP method in (\ref{MPM}) can be applied to obtain an observed neural response sequence $\widehat{\mathcal{R}}_{j}\left(t\right)$. 

Moreover, if $\mathcal{S}\subseteq\mathcal{S}^{\prime}$, then we can also obtain the observed tuning curve of $N_{j}$ as
\begin{equation}
\widehat{\mathcal{G}}_{j}\left(\mathcal{S}^{\prime}\left(m\right)\right)=\widehat{\lambda}_{j}\left(m\right).\label{ObTC}
\end{equation}

\subsection{Summary of neural activity characterization} \label{1.4M}
To this point, we have characterized the stimulus-triggered neural activities in a neural population. For convenience, we define the stochastic process of each neuron $N_{n}$ following
\begin{equation}
\mathcal{P}_{n}^{\heartsuit}\left(r\mid \mathcal{S}^{\prime},0,t\right)=\begin{cases}
\mathcal{P}_{n}\left(r\mid \mathcal{S}^{\prime},0,t\right), & \heartsuit=1 \\
\widehat{\mathcal{P}}_{n}\left(r\mid \mathcal{S}^{\prime},0,t\right), & \heartsuit=0
\end{cases}, \label{MarkerofPP}
\end{equation}
where $\heartsuit$ acts as an index of the type of neuron. $\heartsuit=1$ stands for that $N_{n}$ is an input neuron and its neural activities are defined by (\ref{MasterEq}), while $\heartsuit=0$ means that $N_{n}$ is an intermediary neuron that follows (\ref{MasterEqOb}). 

In our characterization, we distinguish between input and intermediary neurons because input neurons are similar to the sensory cells that perceive external stimuli in the early information processing stage. From a physics perspective, input neurons serve to transform stimuli into non-homogeneous Poisson processes following their neural tuning properties. Although this physics property can be mathematically simplified by direct signal transformation (e.g., see \cite{dayan2001theoretical}), we suggest that the study of neural information function benefits from including input neurons into neural populations.

To computationally generate neural activities in neural populations, one can consider the following algorithm.

\RestyleAlgo{ruled}
\SetKwComment{Comment}{/* }{ */}
\begin{algorithm}[h!]
\caption{Algorithm of neural activity generation}
\KwData{Stimulus sequence $\mathcal{S}^{\prime}$, the number of neurons $m\in\mathbb{N}^{+}$, the proportion of input neurons $\rho\in\left(0,1\right)$, the number of repetition $h$ in (\ref{InterveingIntensityF}), the unit time step $\Delta t$, and the ending time $t^{\prime\prime}$}
\KwResult{Probability distribution and neural response sequence of each neuron $N_{n}$}
\textit{Neural population $\gets$ Randomized following appendix \ref{1.1M}, where there are $m\rho$ input neurons and $m\left(1-\rho\right)$ intermediary neurons}\;
\For{$N_{n}$ in the neural population}{
\If{$N_{n}$ is an input neuron}{
      \textit{Neural tuning curve} $\mathcal{G}_{n}\left(s\right) \gets$ \textit{Randomized following} (\ref{TuningCurve})\;
    }
}
\For{$t \in \{\Delta t,\ldots,t^{\prime\prime}-\Delta t,t^{\prime\prime}\}$}{
\For{$N_{n}$ in the neural population}{
  \eIf{$N_{n}$ is an input neuron}{
    \textit{Probability distribution} $\mathcal{P}_{n}\left(r\mid \mathcal{S}^{\prime},0,t\right) \gets$ \textit{Updated following} (\ref{MasterEq}-\ref{TDF})\;
     \textit{Neural response sequence} $\mathcal{R}_{n}\left(\tau\right)$ \textit{with} $t-\Delta t\leq\tau\leq t \gets$ \textit{Updated following} (\ref{MPM}-\ref{MarkerofNR})\;
  }{\For{$a\in \mathbb{Z}\cap\left[1,h\right]$}{
     \textit{Observed neural response sequence} $\widehat{\mathcal{R}}_{n,a}\left(\tau\right)$ \textit{with} $t-\Delta t\leq\tau\leq t \gets$ \textit{Updated following} (\ref{PostNeuralResponse}-\ref{Threshold})\;}
  \textit{Probability distribution} $\widehat{\mathcal{P}}_{n}\left(r\mid \mathcal{S}^{\prime},0,t\right) \gets$ \textit{Estimated from set} $\{\widehat{\mathcal{R}}_{n,a}\}_{a\in \mathbb{Z}\cap\left[1,h\right]}$ \textit{following} (\ref{InterveingIntensityF}-\ref{MasterEqOb})\;
     \textit{Neural response sequence} $\widehat{\mathcal{R}}_{n}\left(\tau\right)$ \textit{with} $t-\Delta t\leq\tau\leq t \gets$ \textit{Updated following} (\ref{MPM}-\ref{MarkerofNR})\;
  }
}
}

\end{algorithm}

\section{Neural activities as a dynamical system}
\subsection{Neural activities follow the non-homogeneous continuous Markov chain} \label{2.1M}
In (\ref{MasterEq}) and (\ref{MasterEqOb}), we have described the neural activities of both input and intermediary neurons in a neural population with non-homogeneous Poisson processes. A beneficial property for further analysis is that any Poisson process is a kind of continuous Markov chain. Therefore, we reformulate neural activities in the form of the Markov chain.

Assume the neural population encodes a stimulus sequence $\mathcal{S}^{\prime}$ in an interval $\left[0,t^{\prime\prime}\right]$. For a neuron $N_{n}$ in the neural population, we concentrate on the probability of that the neural response rate at moment $t$ is $r$ and that at moment $t+\tau$ ($\tau\geq 0$) is $r^{\prime}$. This probability describes the transformation possibility from neural response state $r$ to state $r^{\prime}$, which can defined as $\upsilon\left(r^{\prime}-r\right)\mathcal{P}_{n}^{\heartsuit}\left(r^{\prime}-r\mid \mathcal{S}^{\prime},t,t+\tau\right)$.

The first step to construct the Markov chain is to define the time-varying transition matrix $\mathcal{W}\left(\mathcal{S}^{\prime},t,t+\tau\right)$ as
\begin{align}
&\mathcal{W}_{ij}\left(\mathcal{S}^{\prime},t,t+\tau\right)=\upsilon\left(j-i\right)\mathcal{P}_{n}^{\heartsuit}\left(j-i\mid\mathcal{S}^{\prime},t,t+\tau\right). \label{TransM}
\end{align}

The second step is to propose the master equation of the Markov chain of neuron $N_{n}$, which is given as
\begin{align}
&\frac{\partial}{\partial t}\mathcal{P}_{n}^{\heartsuit}\left(r\mid \mathcal{S}^{\prime},t,t+\tau\right)=\notag\\&\sum_{r^{\prime}}\Big[\mathbb{W}_{r^{\prime}r}\left(\mathcal{S}^{\prime},t,t+\tau\right)-\mathbb{W}_{rr^{\prime}}\left(\mathcal{S}^{\prime},t,t+\tau\right)\Big],\label{MasterEqofMarkov}
\end{align}
where $\tau\geq0$ and
\begin{align}
\mathbb{W}_{ij}\left(\mathcal{S}^{\prime},t,t+\tau\right)=\mathcal{W}_{ij}\left(\mathcal{S}^{\prime},t,t+\tau\right)\mathcal{P}_{n}^{\heartsuit}\left(i\mid \mathcal{S}^{\prime},0,t\right). 
\end{align}
(\ref{MasterEqofMarkov}) is written in its basic form. Based on (\ref{TransM}), we can also rewrite the master equation as
\begin{align}
&\frac{\partial}{\partial t}\mathcal{P}_{n}^{\heartsuit}\left(r\mid \mathcal{S}^{\prime},t,t+\tau\right)=\notag\\&\sum_{r^{\prime}\leq r}\mathbb{W}_{r^{\prime}r}\left(\mathcal{S}^{\prime},t,t+\tau\right)-\sum_{r^{\prime}> r}\mathbb{W}_{rr^{\prime}}\left(\mathcal{S}^{\prime},t,t+\tau\right). \label{MasterEqofMarkov2}
\end{align}
Based on (\ref{MasterEqofMarkov2}), we can describe the stimulus-triggered neural activity of $N_{n}$ with the non-homogeneous continuous Markov chain.

\subsection{Defining Kolmogorov-Sinai entropy depending on neural tuning properties} \label{2.2M}
Kolmogorov-Sinai entropy characterizes the randomness of a dynamical system forward in time \cite{lecomte2007thermodynamic,gaspard2004time}. The classical definition of the Kolmogorov-Sinai entropy is 
\begin{align}
&\mathcal{H}_{KS}=-\lim_{n\rightarrow +\infty}\frac{1}{n\tau}\notag\\&\sum_{\mathcal{C}_{0},\ldots,\mathcal{C}_{n}}\mathcal{P}\left(\mathcal{C}_{0}\rightarrow\ldots\rightarrow\mathcal{C}_{n}\right)\ln\mathcal{P}\left(\mathcal{C}_{0}\rightarrow\ldots\rightarrow\mathcal{C}_{n}\right), \label{ClassicalDefinitionKS}
\end{align}
where $\tau$ is the variation time step. Each $\mathcal{C}_{i}$ denotes the state of the dynamical system (e.g., the neural response rate we have analysed before). In general, parameter $\tau$ can be understood as the time interval between any two times of sampling for the dynamical system. Enlarging $\tau$ is similar with the coarse graining. One can turn to \cite{lecomte2007thermodynamic,gaspard2004time} for a systematic analysis of Kolmogorov-Sinai entropy in statistical physics.

This research defines the neural activities of each neuron as a kind of non-homogeneous continuous Markov chain. However, this does not mean that we need to consider Kolmogorov-Sinai entropy with a continuous-time limit (the variation time step satisfies $\tau\rightarrow 0$). In the neural system, any kind of variation of neural states takes time (e.g., the time cost of biochemical reaction and the absolute refractory period). Thus, in the calculation of Kolmogorov-Sinai entropy, we only need to consider the situation where the selectable moment (e.g., $t$ in (\ref{MasterEqofMarkov2})) is continuous (this is ensured by the Poisson process) and the variation time step (e.g., $\tau$ in (\ref{MasterEqofMarkov2})) is not approaching to $0$ so as to meet the properties of the neural system. 

Based on the knowledge of Markov chain, it is trivial that (\ref{ClassicalDefinitionKS}) can be written as
\begin{equation}
\mathcal{H}_{KS}=-\frac{1}{\tau}\sum_{\mathcal{C}\mathcal{C}^{\prime}}\mathcal{P}\left(\mathcal{C}\right)\mathcal{W}\left(\mathcal{C}\rightarrow\mathcal{C}^{\prime}\right)\ln\mathcal{W}\left(\mathcal{C}\rightarrow\mathcal{C}^{\prime}\right), \label{ClassicalDefinitionKSM}
\end{equation}
where $\mathcal{P}\left(\mathcal{C}\right)$ is the probability of state $\mathcal{C}$, and $\mathcal{W}\left(\mathcal{C}\rightarrow\mathcal{C}^{\prime}\right)$ defines the transformation probability from state $\mathcal{C}$ to state $\mathcal{C}^{\prime}$ \cite{lecomte2007thermodynamic,gaspard2004time}.

In this research, the probability of neural activities is described in (\ref{MasterEqofMarkov2}). For each neuron $N_{n}$, Kolmogorov-Sinai entropy $\mathcal{H}_{KS}$ can be defined as a metric of its stimulus-triggered neural dynamics (we refer to it as the neural tuning Kolmogorov-Sinai entropy)
\begin{align}
&\mathcal{H}_{KS}\left(N_{n},\mathcal{S}^{\prime},t,t+\tau\right)=-\frac{1}{\tau}\sum_{r}\sum_{r^{\prime}> r}\mathcal{P}_{n}^{\heartsuit}\left(r\mid\mathcal{S}^{\prime},0,t\right)\notag\\
&\times\mathcal{W}_{rr^{\prime}}\left(\mathcal{S}^{\prime},t,t+\tau\right)\ln\mathcal{W}_{rr^{\prime}}\left(\mathcal{S}^{\prime},t,t+\tau\right). \label{NKSE}
\end{align}

Equation (\ref{NKSE}) defines the Kolmogorov-Sinai entropy $\mathcal{H}_{KS}$ of probability distribution $\mathcal{P}_{n}^{\heartsuit}\left(r\mid\mathcal{S}^{\prime},0,t\right)$ and, therefore, proposes a natural approximation of entropy $\mathcal{H}_{KS}$ in the generated neural response trains by $\mathcal{P}_{n}^{\heartsuit}\left(r\mid\mathcal{S}^{\prime},0,t\right)$. The approximation is valid when neural response generation is implemented by the MP method or a large-scale Monte Carlo sampling on $\mathcal{P}_{n}^{\heartsuit}\left(r\mid\mathcal{S}^{\prime},0,t\right)$ \cite{shapiro2003monte} because these two approaches generate neural response trains strictly following $\mathcal{P}_{n}^{\heartsuit}\left(r\mid\mathcal{S}^{\prime},0,t\right)$. However, the approximation becomes invalid when one uses small-scale Monte Carlo sampling or other randomization methods to involve neural response generation with more randomness (e.g., while simulating noisy neural responses). In that case neural response trains do not strictly follow $\mathcal{P}_{n}^{\heartsuit}\left(r\mid\mathcal{S}^{\prime},0,t\right)$ and may have their unique entropy quantities.

\subsection{Chaos of neural activities} \label{2.3M}
An important property of the Kolmogorov-Sinai entropy is that the Kolmogorov-Sinai entropy of a dynamic system is no more than the summation of all the positive Lyapunov exponents of this system (see Ruelle inequality \cite{ruelle1978inequality}). This property connects the Kolmogorov-Sinai entropy with the Lyapunov spectra analysis \cite{engelken2020lyapunov,sompolinsky1988chaos}, supporting an analysis of chaos.

For each neuron $N_{n}$, we have defined its neural tuning Kolmogorov-Sinai entropy by (\ref{NKSE}). Based on Ruelle inequality, there is
\begin{equation}
\mathcal{H}_{KS}\left(N_{n},\mathcal{S}^{\prime},t,t+\tau\right)\leq\sum_{\xi}\mathcal{I}_{\left(0,+\infty\right)}\left(\xi\right)\xi, \label{HKSandLS}
\end{equation} 
where each $\xi$ is a Lyapunov exponent of the dynamic system that describes the neural activities of $N_{n}$, and $\mathcal{I}$ is the indicative function. Here the equality holds only when the system is endowed with an Sinai-Ruelle-Bowen (SRB) measure \cite{ledrappier1985metric}. Under this condition, (\ref{HKSandLS}) is usually referred to as Pesin identity \cite{eckmann1985ergodic}.

In dynamics theory, Lyapunov exponent $\xi$ characterizes the separation or convergence rate of different infinitesimally close trajectories in the phase space (the space of all states of the dynamical system). Here are several key properties of Lyapunov exponent need to be emphasized:
\begin{itemize}
\item For a dynamical system with $n$ parameters, the phase space is $n$-dimensional and there are $n$ Lyapunov exponents;
\item For a Lyapunov exponent $\xi$, if it's positive, then it measures the separation rate of close trajectories in the corresponding direction; if it's negative, then it measures the convergence rate; if it equals $0$, then the trajectories won't separate nor converge in this direction;
\item For a dynamical system with $n$ Lyapunov exponents, if at least one Lyapunov exponent is positive, then the dynamical system can be chaotic.
\end{itemize}

Therefore, if $\mathcal{H}_{KS}$ is positive for neuron $N_{n}$, then the neural activities of this neuron is chaotic.When $\mathcal{H}_{KS}$ increases, the chaos is intensified.

\section{Properties of the neural information processing}
In this section, we will calculate the parameters related to neural encoding and decoding properties. To provide a clear vision, here we give basic explanations for those two conceptions:
\begin{itemize}
\item Neural encoding concerns how neural responses encode and represent the input stimulus;
\item Neural decoding studies how to decode the coded information of stimulus from neural signals.
\end{itemize}
\subsection{Properties of neural encoding} \label{3.1M}
To measure the encoding efficiency of neurons, there are three widely-used parameters:
\begin{itemize}
\item Total response entropy $\mathcal{H}$. It measures the total variation of neural responses to stimuli;
\item Noise entropy $\mathcal{H}^{\vartriangle}$. It measures the variation of neural responses that can not be explained by stimuli;
\item Mutual information $\mathcal{H}^{\vartriangle\vartriangle}$. It measures the variation of neural responses that can be explained by stimuli.
\end{itemize}

The classic definitions of these three parameters have been summarized in \cite{dayan2001theoretical}. In this research, we reformulate the calculation of them to fit dynamic situations. Before implementing the reformulation, there are several necessary derivations to carry out.

First, for the stimulus sequence $\mathcal{S}^{\prime}$ that occurs in $\left[0,t^{\prime\prime}\right]$, it can be recorded to obtain a posteriori probability distribution of stimuli based on the frequency statistics. A stimulus $s$ can not be recorded until it occurs. Thus, the posteriori probability distribution of stimuli is time-dependent (the frequency statistics results of stimuli are updated in real time). For convenience, we use $\mathcal{S}^{\prime}\left(0,t\right)$ to represent the stimulus sequence that has been recorded in $\left[0,t\right]$. For each moment $t$ in $\left[0,t^{\prime\prime}\right]$, the corresponding distribution is defined as $\mathcal{P}\left(\mathcal{S}^{\prime},0,t\right)$. 

Second, for each stimulus $s$, the occurrence duration can be recorded as well. A stimulus might occur multiple times, each time corresponds to an occurrence duration. Although the stimulus sequence $\mathcal{S}^{\prime}$ is sampled from a stationary process in our experiments (therefore, the occurrence duration is fixed as $1$), we still present our theory in a general form to fit more situations. For every moment $t$ in $\left[0,t^{\prime\prime}\right]$, we can do frequency statistics to obtain $\mathcal{P}\left(\tau_{s},0,t\right)$ as the posteriori probability distribution of the occurrence duration length of stimulus $s$ based on the records in $\left[0,t\right]$. Then, we define
\begin{equation}
\tau_{\mathcal{S}}\left(t\right)=\sum_{s\in\mathcal{S}^{\prime}\left(0,t\right)}\mathcal{P}\left(s,0,t\right)\sum_{\tau_{s}}\mathcal{P}\left(\tau_{s},0,t\right)\tau_{s}, \label{EDuration}
\end{equation}
where $\tau_{\mathcal{S}}\left(t\right)$ is the mean duration length averaged from all possible duration length of every stimulus in $\mathcal{S}^{\prime}$ that occurs in $\left[0,t\right]$.

Third, based on the proposed stimulus-triggered neural activity characterization, we can define
\begin{equation}
\mathcal{P}_{n}^{\heartsuit}\left(r,0,t\right)=\sum_{s\in\mathcal{S}^{\prime}\left(0,t\right)}\mathcal{P}_{n}^{\heartsuit}\big(r\mid s,0,\tau_{\mathcal{S}}\left(t\right)\big)\mathcal{P}\left(s,0,t\right),\label{PR}
\end{equation}
where $\mathcal{P}_{n}^{\heartsuit}\left(r\mid s,0,\tau_{\mathcal{S}}\left(t\right)\right)$ denotes the probability for the neural response rate to reach $r$ with a stimulus $s$ lasting for $\tau_{\mathcal{S}}$, which can be calculated by (\ref{MarkerofPP}). It can be seen that (\ref{PR}) defines a neural response probability distribution of $N_{n}$ based on the frequency statistics in $\left[0,t\right]$.

Given these above derivations, we can define the time-dependent total response entropy as
\begin{equation}
\mathcal{H}\left(N_{n},t\right)=-\sum_{r}\mathcal{P}_{n}^{\heartsuit}\left(r,0,t\right)\log_{2}\mathcal{P}_{n}^{\heartsuit}\left(r,0,t\right). \label{TRE}
\end{equation}
Then, we define the time-dependent noise entropy as
\begin{align}
&\mathcal{H}^{\vartriangle}\left(N_{n},t\right)=\sum_{s\in\mathcal{S}^{\prime}\left(0,t\right)}\mathcal{P}\left(s,0,t\right)\notag\\&\times\left[-\sum_{r}\mathcal{P}_{n}^{\heartsuit}\big(r\mid s,0,\tau_{\mathcal{S}}\left(t\right)\big)\log_{2}\mathcal{P}_{n}^{\heartsuit}\big(r\mid s,0,\tau_{\mathcal{S}}\left(t\right)\big)\right]. \label{NE}
\end{align}
Finally, we can calculate time-dependent mutual information as
\begin{equation}
\mathcal{H}^{\vartriangle\vartriangle}\left(N_{n},t\right)=\mathcal{H}\left(N_{n},t\right)-\mathcal{H}^{\vartriangle}\left(N_{n},t\right). \label{MI}
\end{equation}

Moreover, we can also measure the time-dependent entropy of the stimulus sequence as
\begin{equation}
\mathcal{H}_{\mathcal{S}}\left(\mathcal{S}^{\prime},t\right)=-\sum_{s\in\mathcal{S}^{\prime}\left(0,t\right)}\mathcal{P}\left(s,0,t\right)\log_{2}\mathcal{P}\left(s,0,t\right). \label{EofS}
\end{equation}
If we further define
\begin{align}
    &\mathcal{H}_{s}^{\vartriangle}\left(N_{n},t\right)=\notag\\&-\sum_{r}\mathcal{P}_{n}^{\heartsuit}\big(r\mid s,0,\tau_{\mathcal{S}}\left(t\right)\big)\log_{2}\mathcal{P}_{n}^{\heartsuit}\big(r\mid s,0,\tau_{\mathcal{S}}\left(t\right)\big), \label{NEofS}
\end{align}
we can measure the noise from the encoding of stimulus $s$. It is clear that the noise entropy $\mathcal{H}^{\vartriangle}$ is averaged from $\mathcal{H}_{s}^{\vartriangle}\left(N_{n},t\right)$ of every stimulus. For each stimulus $s$, if $\mathcal{H}_{s}^{\vartriangle}\left(N_{n},t\right)<\mathcal{H}^{\vartriangle}\left(N_{n},t\right)$, then the noise from the encoding process of it is relatively less. In other words, the interpretability of neural activities based on $s$ is relatively high. Given this definition, if we define a specific subset
\begin{equation}
\mathcal{S}_{\mu}\left(0,t\right)=\lbrace s\mid \mathcal{H}_{s}^{\vartriangle}\left(N_{n},t\right)<\mathcal{H}^{\vartriangle}\left(N_{n},t\right)\rbrace, \label{CSBases}
\end{equation}
and define that
\begin{equation}
\mu\left(N_{n},t\right)=\dfrac{\vert\mathcal{S}_{\mu}\left(0,t\right)\vert}{\vert\mathcal{S}^{\prime}\left(0,t\right)\vert}, \label{CodingS}
\end{equation}
then $\mu\left(N_{n},t\right)$ measures the proportion of the stimuli that can better explain neural activities in all stimuli. We refer to $\mu$ as the encoding scope.

Till now, we can calculate the total response entropy, noise entropy, and mutual information only based on the stimulus in $\mathcal{S}_{\mu}$ (here $\mathcal{S}_{\mu}$ can be treated as specific local stimulus distribution). Most parts of calculations keep the same with what has been defined above. The only two differences lie in that we need to recalculate the occurrence probability of the stimuli in $\mathcal{S}_{\mu}$ and the occurrence duration length (since we only focus on the local part of stimulus distribution). Specifically, for each $s$ in $\mathcal{S}_{\mu}$, its new probability $\mathcal{P}^{\prime}\left(s,0,t\right)$ is recalculated as
\begin{equation}
\mathcal{P}^{\prime}\left(s,0,t\right)=\frac{\mathcal{P}\left(s,0,t\right)}{\sum_{s_{i}\in\mathcal{S}_{\mu}}\mathcal{P}^{\prime}\left(s_{i},0,t\right)}. \label{RecalculationP}
\end{equation}
Apart from that, the new occurrence duration length is
 \begin{equation}
\tau_{\mathcal{S}}^{\prime}\left(t\right)=\sum_{s\in\mathcal{S}_{\mu}\left(0,t\right)}\mathcal{P}^{\prime}\left(s,0,t\right)\sum_{\tau_{s}}\mathcal{P}\left(\tau_{s},0,t\right)\tau_{s}. \label{RecalculationDur}
\end{equation}
Based on the new probability redefined in (\ref{RecalculationP}-\ref{RecalculationDur}), we can further calculate all parameters based on (\ref{PR}-\ref{MI}).

\subsection{Properties of neural decoding}\label{3.2M}
Decoding the stimulus parameter from given neural signals is important for neuroscience studies. With a decoding (or estimation) technique, researchers can predict the input stimulus based on the neural response. 

For neuron $N_{n}$, assume that we have recorded its neural response sequence to $\mathcal{S}^{\prime}$ in $\left[0,t^{\prime\prime}\right]$. At any moment $t$ in $\left[0,t^{\prime\prime}\right]$, a decoding technique is applied to obtain a time-dependent estimated stimulus sequence $\mathcal{S}_{est}^{\prime}\left(0,t\right)$ based on the neural response sequence in $\left[0,t\right]$. If we repeat the experiment with same stimulus sequence $\mathcal{S}^{\prime}$ for $k$ times and each time we do an estimation, then we can obtain an averaged time-dependent estimate
\begin{equation}
\forall s\in\mathcal{S}^{\prime}\left(0,t\right),\; \langle s_{est}\left(0,t\right)\rangle=\frac{1}{k}\sum_{i=1}^{k}s_{est}\left(0,t,i\right)
\end{equation}
where $s_{est}\left(0,t,i\right)$ is the estimated result of stimulus $s$ of $\mathcal{S}^{\prime}\left(0,t\right)$ in the $i$th experiment.

 Then, we can measure the time-dependent accuracy of the decoding technique applied on neuron $N_{n}$ with the bias $\mathcal{B}_{n}\left(t\right)$ and the variance $\mathcal{D}_{n}\left(t\right)$ as
\begin{equation}
\mathcal{B}_{n}\left(t\right)=\sum_{s\in\mathcal{S}^{\prime}\left(0,t\right)}\mathcal{P}\left(s,0,t\right)\Big(s-\langle s_{est}\left(0,t\right)\rangle\Big),
\end{equation}
and 
\begin{align}
&\mathcal{D}_{n}\left(t\right)=\notag\\&\frac{1}{k}\sum_{i=1}^{k}\left[\sum_{s\in\mathcal{S}^{\prime}\left(0,t\right)}\mathcal{P}\left(s_{est},0,t\right)\Big(s_{est}\left(0,t,i\right)-\langle s_{est}\left(0,t\right)\rangle\Big)^2\right].
\end{align}
For a decoding technique, it is optimized if its variance approaches to $0$. 

In statistical theory, the Cram$\acute{\text{e}}$r-Rao bound suggests that the Fisher information limits the accuracy with which any decoding technique can estimate the target parameter of the stimulus. Assume that $\mathcal{B}^{\mathscr{E}}_{n}\left(t\right)$ and $\mathcal{D}^{\mathscr{E}}_{n}\left(t\right)$ are the bias and variance obtained based on a decoding technique $\mathscr{E}$ that is applied on $N_{n}$. Then the Cram$\acute{\text{e}}$r-Rao bound can be given as \cite{dayan2001theoretical}
\begin{equation}
\forall \mathscr{E},\;\mathcal{D}^{\mathscr{E}}_{n}\left(t\right)\geq\dfrac{\left(1+\frac{\partial}{\partial t}\mathcal{B}^{\mathscr{E}}_{n}\left(t\right)\right)^2}{\mathcal{F}\left(N_{n},t\right)}, \label{CRBound}
\end{equation}
where $\mathcal{F}\left(N_{n},t\right)$ denotes the time-dependent Fisher information of neuron $N_{n}$. Based on (\ref{CRBound}), it can be seen that the calculation of $\mathcal{F}\left(N_{n},t\right)$ is important since it acts as the lower bound of the variance of any decoding technique applied on neuron $N_{n}$. Even for the unbiased decoding ($\mathcal{B}^{\mathscr{E}}_{n}\left(t\right)\equiv0$, thus $\frac{\partial}{\partial t}\mathcal{B}^{\mathscr{E}}_{n}\left(t\right)\equiv0$), the variance is still no less than $\frac{1}{\mathcal{F}\left(N_{n},t\right)}$ \cite{dayan2001theoretical}. 

To measure the precision limitation of any possible decoding technique that can be applied on neuron $N_{n}$, we calculate the time-dependent Fisher information $\mathcal{F}\left(N_{n},t\right)$ as
\begin{align}
&\mathcal{F}\left(N_{n},t\right)=\sum_{s\in\mathcal{S}^{\prime}\left(0,t\right)}\mathcal{P}\left(s,0,t\right)\notag\\&\left\{\sum_{r}\mathcal{P}_{n}^{\heartsuit}\big(r\mid s,0,\tau_{\mathcal{S}}\left(t\right)\big)\left[\Delta_{s}\Big(\ln \mathcal{P}_{n}^{\heartsuit}\big(r\mid s,0,\tau_{\mathcal{S}}\left(t\right)\big)\Big)\right]^2\right\}, \label{FisherInformation}
\end{align}
where $\Delta_{s}\left(\cdot\right)$ is the first order difference with respect to $s$. The Fisher information proposed in (\ref{FisherInformation}) is in discrete form. In special cases, if $\mathcal{P}_{n}^{\heartsuit}\big(r\mid s,0,\tau_{\mathcal{S}}\left(t\right)\big)$ is sufficiently smooth with respect to $s$, we can also use the partial derivative to replace the first order difference to obtain a continuous form. Moreover, the calculated quantity in (\ref{FisherInformation}) is the Fisher information of the whole stimulus sequence. It can be treated as the expectation of the Fisher information $\left\{\sum_{r}\mathcal{P}_{n}^{\heartsuit}\big(r\mid s,0,\tau_{\mathcal{S}}\left(t\right)\big)\left[\Delta_{s}\Big(\ln \mathcal{P}_{n}^{\heartsuit}\big(r\mid s,0,\tau_{\mathcal{S}}\left(t\right)\big)\Big)\right]^2\right\}$ of each stimulus $s$ in the sequence.

For a decoding method $\mathscr{E}$, it is optimal if and only if its variance satisfies $\mathcal{D}^{\mathscr{E}}_{n}\left(t\right)=\frac{1}{\mathcal{F}\left(N_{n},t\right)}$. And it can be seen that for neuron $N_{n}$, if its Fisher information increases, then the optimal decoding technique applied on it can realize a better precision.

To this point, we have defined the information processing properties of neural activities from the perspectives of encoding and decoding. Similar to the neural tuning Kolmogorov-Sinai entropy $\mathcal{H}_{KS}$, those information processing properties are defined based on probability distribution $\mathcal{P}_{n}^{\heartsuit}\left(r\mid\mathcal{S}^{\prime},0,t\right)$, approximating the properties of neural response trains. The validity of approximation relays on whether neural responses strictly follow $\mathcal{P}_{n}^{\heartsuit}\left(r\mid\mathcal{S}^{\prime},0,t\right)$ and, consequently, can not be ensured when one generate noisy neural neural responses with more randomness.
 \end{appendix}


\bibliography{apssamp}
\end{document}